# Transforming U.S. agriculture with crushed rock for $CO_2$ sequestration and increased production

David J. Beerling[1]*, Euripides P. Kantzas[1], Maria Val Martin[1], Mark R. Lomas[1], Lyla L. Taylor[1], Shuang Zhang[2], Yoshiki Kanzaki[3], Christopher T. Reinhard[3], Noah J. Planavsky[4], Rafael M. Eufrasio[5], Phil Renforth[6], Jean-Francois Mecure[7,8], Hector Pollitt[7,8], Philip B. Holden[9], Neil R. Edwards[9], Lenny Koh[5], Dimitar Z. Epihov[1], Adam Wolf[10], James E. Hansen[11], Nick F. Pidgeon[12] & Steven A. Banwart[13,14]

[1]Leverhulme Centre for Climate Change Mitigation, School of Biosciences, University of Sheffield, Sheffield, UK

[2]Department of Oceanography, Texas A&M University, College Station, TX

[3]School of Earth and Atmospheric Sciences, Georgia Institute of Technology, Atlanta, USA

[4]Department of Earth and Planetary Sciences, Yale University, New Haven, USA

[5]Advanced Resource Efficiency Centre, Management School, University of Sheffield, Sheffield, UK

[6]School of Engineering and Physical Sciences, Heriot-Watt University, Edinburgh Campus, Edinburgh, UK

[7]Global Systems Institute, Department of Geography, University of Exeter, Exeter, UK

[8]Cambridge Centre for Energy, Environment and Natural Resource Governance, University of Cambridge, Cambridge, UK

[9]Environment, Earth and Ecosystems, The Open University, Milton Keynes, UK

[10]Department of Ecology and Evolutionary Biology, Princeton University, Princeton, USA

[11]Earth Institute, Columbia University, New York, NY, USA

[12]Understanding Risk Research Group, School of Psychology, Cardiff University, Cardiff, UK

[13]Global Food and Environment Institute, University of Leeds, Leeds, UK

[14]School of Earth and Environment, University of Leeds, Leeds, UK

*Corresponding author. Email: d.j.beerling@sheffield.ac.uk

**Author contributions.** Conceptualization: DJB, EK, MRL, MvM, SAB, LLT, NJP, CTR. Methodology: EPK, MRL, MvM, LLT, SZ, YK, CTR, DJB. Funding acquisition: DJB, MvM, NJP. Writing – original draft: DJB, SAB, MvM, NFP, SZ, CTR, NJP, AW, JEH. Writing – review & editing: all authors.

**Competing interests:** D.J.B. has a minority equity stake in Future Forest/Undo. Adam Wolf is financially supported by Eion Corp, A Public Benefit Corporation, where he is also a minority shareholder. All other authors declare that they have no competing interests.

Classification: Biological Sciences.

**This PDF file includes:**
Main Text
Figures 1 to 5

**Abstract**

**Enhanced weathering (EW) is a promising modification to current agricultural practices that uses crushed silicate rocks to drive carbon dioxide removal (CDR). If widely adopted on farmlands, it could help achieve net-zero or negative emissions by 2050. We report detailed state-level analysis indicating EW deployed on agricultural land could sequester 0.23-0.38 Gt $CO_2$ $yr^{-1}$ and meet 36-60 % of U.S. technological CDR goals. Average CDR costs vary between state, being highest in the first decades before declining to a range of ~$100-150 $tCO_2^{-1}$ by 2050, including for three states (Iowa, Illinois, and Indiana) that contribute most to total national CDR. We identify multiple electoral swing states as being essential for scaling EW that are also key beneficiaries of the practice, indicating the need for strong bipartisan support of this technology. Assessment the geochemical capacity of rivers and oceans to carry dissolved EW products from soil drainage suggests EW provides secure long-term $CO_2$ removal on intergenerational time scales. We additionally forecast mitigation of ground-level ozone increases expected with future climate change, as an indirect benefit of EW, and consequent avoidance of yield reductions. Our assessment supports EW as a practical innovation for leveraging agriculture to enable positive action on climate change with adherence to federal environmental justice priorities. However, implementing a stage-gating framework as upscaling proceeds to safeguard against environmental and biodiversity concerns will be essential.**

**Significance statement**

Given evidence that current greenhouse gas concentrations are already having substantial economic and health impacts, scalable atmospheric carbon dioxide removal (CDR) strategies will be needed for mitigating future climate change. Enhanced weathering (EW) is emerging as one of the few CDR technologies that could achieve a billion tonnes of $CO_2$ removal within decades. However, delivering on this milestone requires national strategies for deployment. Our detailed initial assessment for the U.S. suggests EW deployed on croplands in the contiguous U.S. could sequester 0.23-0.38 Gt $CO_2$ $yr^{-1}$ to meet 36-60 % of nation's technological CDR goals. Upscaling EW has scope to create a new multi-billion-dollar CDR industry that affords potential opportunities for environmental justice in decarbonisation but requires environmental safeguards. Revenue generation by EW can contribute to market opportunities, new employment options, and rural economic diversification.

**Introduction**

The United States (U.S.) strategy to reach net-zero greenhouse gas (GHG) emissions by 2050 includes decarbonizing the energy system and deployment of carbon dioxide removal (CDR) technologies at scale (1-3) to sequester a billion metric tons of $CO_2$ annually (1 Gt $CO_2$ $yr^{-1}$) within three decades. CDR strategies are needed to achieve a net-zero carbon budget which is otherwise unlikely given "hard to decarbonize industries" such as agriculture, cement production, and aviation. The U.S. CDR goal is to be achieved by enhancing natural land carbon sinks and scaling up $CO_2$ removal technologies, mainly Bioenergy with Carbon Capture and Storage (BECCS) and Direct Air Carbon Capture and Storage (DACCS) to deliver 0.5 Gt CDR (1). Given evidence that current GHG concentrations are already well into the dangerous level (4), there is a possibility that even greater CDR will be needed to achieve a negative carbon budget over the coming century.

Here we focus on the potential of purposeful enhanced rock weathering (EW) in agricultural settings as a promising but still underexplored CDR technology for meeting U.S. decarbonization targets (1-3). EW is an *ex-situ* enhanced mineralization approach (5, 6). Most commonly, cropland soils are amended with crushed basalt to drive $CO_2$ removal, improve food and soil security (7, 8). An additional benefit of EW being mitigation of ocean acidification (9-11). The approach has lower intrinsic costs than either BECCS or DACCS at the outset (3, 6). Moreover, EW can leverage existing transportation infrastructure and technology for rapid scalability with significant social and environmental benefits while supporting rural economic development, a key U.S. national policy priority (12)

Global biogeochemical modeling of the CDR potential of EW on agricultural lands indicates that the U.S. has overwhelming potential to lead on the implementation of this technology (6). The Corn Belt in the American Midwest alone has over 70 million hectares in corn and soybean rotation and represents one of the most intensively managed agricultural regions in the world—providing obvious opportunities for integrating EW practices. Large-scale replicated EW field trials with crushed basalt in the heart of the U.S. Corn Belt provide strong support for deployment, demonstrating carbon sequestration and major agronomic benefits, including enhanced yields of key food and feed crops via improved soil fertility (13). However, as with other CDR strategies, EW is at a nascent stage of research, development, and demonstration (14) and the U.S. Congress has prioritized the need to quantify costs, risks, and benefits of EW to set the stage for widespread implementation (15).

We present an integrated whole-system assessment of EW as a CDR technology rapidly deployable with U.S. agriculture to mitigate climate impacts while creating a new multi-billion-dollar industry. State-specific low- and high-basalt supply rates, via quarrying, provide feedstock constraints for high-resolution dynamic climate-nitrogen-carbon cycle simulations (16) of EW in soils and CDR between 2020 and 2070 (figs. S1-S7). We quantify CDR and security of $CO_2$ removal by assessing the geochemical capacity of rivers and oceans to carry dissolved EW products from soil drainage (Methods figs. S8-S12). Technological development pathways for EW are constrained by both basalt supply rate and regional future energy policies consistent with 1.5°C of warming by 2050 (17, 18)(figs. S13-S15). We further quantify benefits for soils and regional air-quality and assess state-level CDR economics and discuss the social license to operate this technology.

**Results and Discussion**

**Carbon drawdown potential of U.S. agriculture**

Results indicate a net CDR potential for U.S. agriculture of between 0.23 Gt $CO_2$ $yr^{-1}$ and 0.38 Gt $CO_2$ $yr^{-1}$ by 2050 for our low and high rock-extraction scenarios, respectively, with relatively f-grained basalt (i.e., 80% of particles less than or equal to 100 µm diameter) (**Fig. 1A**). These rates increase further to 0.25 Gt $CO_2$ $yr^{-1}$ and 0.48 Gt $CO_2$ $yr^{-1}$ by 2070, equivalent to ~6% of current U.S. emissions. By 2060, 85-90% of the total CDR in the U.S. for both rock extraction scenarios is accounted for by ten of the twenty states analysed, with the CDR of each rising over successive decades following repeated annual crushed rock applications (**Fig. 1B**). Of those ten states, four Corn Belt states (Illinois, Iowa, Indiana, Missouri) make the largest contributions, in part, by virtue of having a large geographical area for rock dust deployment reaching net 40-75 Mt $CO_2$ removal $yr^{-1}$ by 2050 (**Fig. 1B**). Three other Corn Belt states (Wisconsin, Minnesota, Michigan) are early EW adopters but are smaller in area, reaching average CDR rates of 15-20 Mt $CO_2$ $yr^{-1}$ by 2050 (**Fig. 1B**).

The trajectory of CDR by EW tracks cumulative rock supply for basalt quarrying states projected into the future using a sigmodal function that describes a slow initial roll-out phase of 10-20 years (with delays due to quarrying licenses, public acceptance), followed by rapid expansion with the opening of new quarries, before finally reaching steady-state production rates. Although supply-side dynamics, as constrained by state-specific historical data, are somewhat uncertain, our framework for this initial assessment gives dynamics compatible with the trajectory of CDR strategies adopted by integrated assessment models in pathways designed to reach net-zero by 2050 (19, 20).

Geographical patterns of CDR rates per hectare (**Fig. 1 C,D**) reflect timing when EW is initiated, which is a function of proximity to basalt supply, soil pH, crop type, and climate. We assume supply states nearest to the agricultural states respond quickest and that those states with existing infrastructure for rock extraction are best placed to achieve the highest production rates. Earlier EW deployment allows greater cumulative CDR due to consecutive annual applications, with basalt added in earlier years still capturing $CO_2$ as the slow-weathering minerals dissolve, as observed in EW field trials (13). Corn Belt states closest to basalt production states start EW early and have the highest CDR rates per hectare potential by 2070 (**Fig. 1D**). Optimal selection of states for basalt provision to croplands is based on minimizing transportation distances, thus maximising CDR by keeping logistical $CO_2$ emissions low (which also reduces overall cost) and delivering EW efficiently over time. This analysis identifies states to prioritize for early EW implementation to maximize CDR potential.

Overall, our results suggest EW deployment with U.S. agriculture could contribute 36-60 % of U.S. technological CDR goals (1) and 18-30% of overall CDR goals by 2050. This represents a substantial contribution to near-term U.S. net-zero pathways. Relative to both engineered and other terrestrial CDR options proposed to achieve net-zero, potential rates of $CO_2$ removal by EW are competitive and warrant consideration for large-scale implementation. BECCS, for example, has an estimated U.S. technical potential 0.36-0.63 Gt $CO_2$ $yr^{-1}$ in 2040, after accounting for constraints of long-distance biomass and $CO_2$ transport, regional $CO_2$ storage, and injection well capacities (21). Afforestation/reforestation (0.25-0.6 Gt $CO_2$ $yr^{-1}$), agricultural practices to increase soil carbon sequestration (0.25 Gt $CO_2$ $yr^{-1}$) and reforestation of understocked timberlands (0.19 Gt $CO_2$ $yr^{-1}$) (22), all have similar CDR potential in the U.S. to that modelled here for EW (0.18-0.38 Gt $CO_2$ $yr^{-1}$ by 2050) (14), but questions exist about the permanence and storage capacity of these practices.

Analysis of crushed rock transfer between source states and recipient agricultural states for achieving these CDR trajectories shows that within a couple of decades three states with pre-existing quarrying infrastructure co-located with basalt reserves (Wisconsin, Minnesota and Michigan) are the main rock suppliers to adjacent farmland (2040) (**Fig. 1E**). By 2070, basalt supply for EW ramps up to include seven key states meeting the demand of eleven main crop states (**Fig. 1F**), with Virginia, North Carolina and Pennsylvania becoming additionally important.

Implementation of EW at scale requires interstate transportation networks with sufficient capacity for moving basalt from supply states to crop states as well as engagement of multiple stakeholders for producing, collecting, and transporting crushed rock. However, cost-effective quarrying and transporting of material at scale are activities society undertakes today, with technology, infrastructure, and human capital already in place. In contrast, BECCS and DACCS require the scaling up of novel technologies and supporting infrastructure and retraining of an existing workforce. Given the use of existing infrastructure and our process-based approach, we consider our modelled CDR rates to be realisable, assuming financial support is forthcoming.

Rock grinding is the dominant energy demanding step in the EW supply chain in our at scale simulations (14, 16, 23). Grinding requires 0.4% (20 TWh) and 0.2% (1.5 TWh) of electricity production of the eastern and western power grids, respectively, for undertaking EW with up to 1 Gt rock yr$^{-1}$ by 2070 (fig. S16). These figures rise to 0.8% and 0.4% by 2070 for EW with the 2 Gt rock yr$^{-1}$ (fig. S16) and fall at or below the range of current national power usage for rock comminution processes (USA 0.4%, Canada 1.9%, South Africa 1.8% and Australia 1.5%) (24), countering concerns regarding excessive energy demand from upscaling EW.

**Durability of carbon sequestration**

We next quantify durability of CDR with EW by assessing the capacity of U.S. rivers to carry dissolved EW products from soil drainage without re-release of $CO_2$ via carbonate precipitation (fig. S9-S12) (21). Our analysis uses six major U.S. river watershed water chemistry and flow data for 863 river sites (**Fig. 2 A,B**), updated with soil water drainage weathering products and climate model data, to calculate changes in the flow and carbonate saturation state ($\Omega$) of river systems, 2020-2070 (25).

Carbonate precipitation and $CO_2$ re-release in river systems is likely negligible at $\Omega$ below 10 and accordingly the distribution of current $\Omega$ values in U.S. rivers indicates negligible precipitation at the catchment scale (**Fig. 2 C,D**) (25). Simulation results show that with excess amounts of solutes (e.g., Ca, Mg, Na, K, $HCO_3^-$) derived from EW entering river systems, $\Omega$ remains generally below the kinetic threshold required for extensive carbonate precipitation, i.e., $\Omega$ less than 10, for the 1 and 2 Gt rock yr$^{-1}$ extraction scenarios (**Fig. 2D**) (21). River $\Omega$ values increase with time due to increasing EW solute fluxes, but more than 86% of rivers across both scenarios have $\Omega$ values less than 10 by 2060-2070 (**Fig. 2 C,D**), indicating the majority have capacity to transport weathered products without $CO_2$ re-release. Additionally, there are likely be high rates of carbonate dissolution in the upper portion of riverine sediments - even in limited cases of water column carbon formation (26). Therefore, our river geochemistry calculations indicate that transport of dissolved constituents in surface waters is unlikely to be a bottleneck limiting the CDR potential of EW (25).

We calculate potential for $CO_2$ leakage following the transportation of the EW products by rivers to the ocean with a 3-D ocean biogeochemistry model at locations representing the outlets of the six major watersheds (**Fig. 2B**). Results indicate a compensatory ocean outgassing of $CO_2$ with EW deployment due to equilibration of the ocean-atmosphere system that gradually increases over time to around 10% in 2040 and 25% in 2070 (**Fig. 2 E,F**) (10, 16, 27). This is a well-established Earth system response that occurs with all CDR technologies (27). However, we also calculate a further ~5% backflow of $CO_2$ out of the ocean caused by the re-equilibration of the shallow ocean carbonate system that represents carbon captured by EW returning to the atmosphere on short timescales, which represents an inefficiency in EW (**Fig. 2 E,F**).

**Soil pH and nutrient responses to EW**

Reversing widespread long-term acidification of U.S. agricultural soils driven by inorganic fertilizer usage (28) is important for improving crop yields, soil fertility, and nitrogen fixation rates in legumes (e.g., soybean) (29). We show that EW addresses this situation by progressively increasing average farmland soil pH from 6.4 in 2020 to pH 6.7 or 7.1 in 2070 with the 1 and 2 Gt $yr^{-1}$ rock extraction scenarios, respectively (**Fig. 3A**). These results support substituting basalt for agricultural limestone to manage soil acidity, given that limestone used in U.S. agriculture releases more than 5 Mt $CO_2$ annually when applied to low pH soils (30). Average topsoil pH typically remains close to the optimal range of nutrient uptake by major row crops even after decades of continuous EW implementation (7) (fig. S17). Agriculture in the western U.S. may experience soil pH increases above 7.5 (fig. S17), where EW could be stopped to avoid micronutrient deficiency without major impact on overall U.S. CDR (<5%). Over time, EW reduces the fraction of soils classified as acidic (i.e., pH less 6.5) from ~0.58 to between 0.3 and 0.05 by 2070, when acidified soils are practically eliminated (**Fig. 3B**).

EW releases crop nutrients phosphorus (P) and potassium (K), as basaltic minerals undergo dissolution, to reduce the requirement for P and K nutrients that are now usually obtained from expensive chemical fertilizers with large environmental footprints and geopolitically unstable global supply chains. Calculated P release patterns by EW reflect fast-apatite weathering and across the top ten Corn Belt states (by CDR potential) release rates are typically 10-30 kg $ha^{-1}$ $yr^{-1}$ (**Fig. 3C**). For K release from slower-weathering feldspars, rates increase from 20 kg $ha^{-1}$ $yr^{-1}$ in 2030-2040 to 60-70 kg $ha^{-1}$ $yr^{-1}$ by 2070 (**Fig. 3D**). These rates are comparable to the range of maintenance P and K fertilizer application rates used in the Mid-west for soybean, maize and wheat (**Fig. 3 C,D**), although rates vary with crop and soil type. Rock supply states for farmland in the western U.S. produce basalt with a higher P and K content than on the East coast, and EW nutrient release rates that could exceed current average fertilizer rates (figs. S3, S18, S19). Based on current agronomic practices for three crops (soybean, wheat and maize) across ten states, our analysis reveals EW practices can partially or completely replace expensive P (urea phosphate, $890 $t^{-1}$; diammonium phosphate (DAP), $938 $t^{-1}$) and K (potash $862 $t^{-1}$) (31) fertilizers, thereby avoiding millions of tons of $CO_2$ emissions linked to their production and distribution (16).

**Air quality improvements with EW deployment**

Mitigation of nitrous oxide ($N_2O$) emissions from agriculture, as the dominant anthropogenic source of $N_2O$, is an integral part of U.S. net-zero pathways because $N_2O$ is a potent long-lived greenhouse gas (19) that also causes stratospheric ozone depletion (32). It is also an important demonstrable benefit of EW with farmland (33). We find $N_2O$ emissions with EW are reduced by 20-30% in 2050 and 25-40% in 2070 compared to the control case without EW and expected future N-fertilizer usage (fig. S20). Expressed as $CO_2$ equivalents, the EW $N_2O$ reductions translate into 90 Mt $CO_{2,e}$ $yr^{-1}$ and ~120 Mt $CO_{2,e}$ $yr^{-1}$ of avoided emissions by 2070 (1 Gt and 2 Gt rock $yr^{-1}$ scenarios, respectively, fig. S20) and improve EW greenhouse gas removal budgets by a further 36-45%. In comparison, other $N_2O$ mitigation approaches, such as changing fertilizer management practices to increase the efficiency of plant nitrogen uptake (1), are less effective than EW (mitigating 8.8 $MtCO_{2,e}$ $yr^{-1}$ by 2030).

Air quality improvements follow because soil emissions of nitric oxide (NO) track decreases in $N_2O$ emissions, with reductions of 15-25% and 25-30% by 2050 for the 1 and 2 Gt $yr^{-1}$ rock extraction scenarios, respectively (fig. S20). These reductions occur due to the rise in soil pH with EW increasing the ratio of $N_2$ to $N_2O$ production during denitrification (34). Soil NO released to the atmosphere undergoes rapid oxidation to nitrogen dioxide ($NO_2$) to generate tropospheric ozone ($O_3$) (35), a strong oxidant detrimental to crop health (36). Thus, lowering NO emissions with EW can decrease ground-level $O_3$ production (fig. S21). We focus on the response of AOT40 values (defined as Accumulated dose of Ozone over a Threshold of 40 ppb), an ozone indicator for protecting crop health. Decade-mean AOT40 values are dominated by anthropogenic emissions and tend to be highest over highly populated and industrial regions, such as California and the Eastern U.S. (**Fig. 4A**). With EW, however, we find deep and widespread reductions in AOT40 throughout the Corn Belt in 2050, which expand further by 2070 (**Fig. 4A**). Averaged by state, AOT40 reductions are 2-4 ppm-hr (**Fig. 4B**) and translate into yield gains of 2-4% for maize and soybean, and up to 6% for wheat (**Fig. 4C**). These gains are comparable to those obtained for crops with specific air quality improvement scenarios (37). Mitigation of ground-level $O_3$, an overlooked indirect benefit of EW, is sufficient to counteract increases expected with future climate change (38). We calculate that the damage-avoided value for ozone-induced yield losses of individual states could be worth $100-$500 million annually by 2070 (**Fig. 4D**).

Increases in soil pH with EW carry the risk of increasing aerosol pollutants harmful to human health by stimulating ammonia ($NH_3$) volatilization at higher pH (39). Ammonia plays an important role in the formation of fine particulate matter (particles < 2.5µm in diameter; $PM_{2.5}$). It reacts with nitric acid, derived from the reaction of $NO_2$ with water, to form secondary inorganic nitrate and ammonium aerosols that catalyze production of $PM_{2.5}$. In our modelling EW increases soil $NH_3$ emissions by 4-6% by 2070 (fig. S22), but with reduced soil NO emissions limiting production of nitric acid, the formation of secondary inorganic nitrate and ammonium aerosols is reduced. Consequently, we calculate 5% reduction in spring and summertime $PM_{2.5}$ by 2070 (fig. S22). This suggests EW may be an effective measure to control future $PM_{2.5}$ formation in agricultural regions and is consistent with reducing emissions of nitrogen oxides to control $PM_{2.5}$ in California (40). Nonetheless, rock dust emitted directly during basalt applications may be an important source of primary aerosols, which can temporarily increase local $PM_{2.5}$ levels and require safety procedures to minimize human exposure (e.g., positive pressure tractor cabs, respirators at transfer sites) (41).

**Costs and revenue generation by EW**

Geospatially mapped CDR costs show marked heterogeneity across the U.S. reflecting a combination of cropland distance from basalt source regions, timing of EW deployment and evolving CDR rates (**Fig. 5 A,B**). By 2040-2050, spatial patterns of CDR costs are lowest in crop states that start EW deployment early, such as Minnesota and Wisconsin (**Fig. 5A**). By 2060-2070, widespread reductions in CDR costs occur due to cumulative increases in CDR and lower energy costs for transportation and rock grinding, with most states achieving CDR at a cost of ≤ $150 $tCO_2^{-1}$ (**Fig. 5B**). Central U.S. agricultural states (e.g., Kansas and Nebraska) are the exception where costs remain stubbornly high mainly due to long transport distances of crushed rock from supply states; however, these states are relatively minor contributors to total U.S. CDR

(**Fig. 1B**). This geospatial cost analysis strengthens support for early implementation of EW in northerly Corn Belt states.

Average CDR costs vary between state, being highest in the first decades before declining to a range of ~$100-130 tCO$_2^{-1}$ by 2050, including three states (Iowa, Illinois, and Indiana) that contribute most to the U.S. total (**Fig. 5C**). This range compares favourably with an initial global technoeconomic assessment of EW for 2050 that included estimates for the U.S. ($160-180 tCO$_2^{-1}$) (*5*). Our analysis indicates that most agricultural states undertaking EW reach the suggested threshold (~$100 tCO$_2^{-1}$) for making CDR technologies affordable and ready for large-scale immediate deployment (*14*) within a decade or two.

Opportunities exist for marked cost reductions in CDR with EW. Long transport distances between basalt and crop states in the U.S. compared to e.g., the United Kingdom (*16*), are a key factor contributing to variations in CDR costs between states by 2050 (**Fig. 5D**). Mass transportation of crushed rock from source states to crop states by high-capacity barge traffic on navigable inland river systems could help to lower these costs and additionally reduce operational $CO_2$ emissions; barges currently emit 25% less $CO_2$ than railroads per ton mile (*42*).

Further, long-term EW field trials in the Corn Belt show yield increases for maize and soybean of 12% and 16%, respectively, after 4 years of treatment (*13*). Combining state-level economic gains per hectare resulting from maize and soybean yield increases, with corresponding average CDR rates per hectare, gives a 'yield discount' on CDR cost per ton of $CO_2$. This discount is highest in the first couple of decades of EW deployment typically ~$40-60 tCO$_2^{-1}$ prior to CDR rates ramping up over time (**Fig. 5E**). Consequently, 'yield discounts' fortuitously favour lowering CDR costs early in the deployment phase when initial costs are highest (**Fig. 5C**).

Calculated costs of CDR with EW are, however, already competitive compared to other durable engineered CDR strategies. These other CDR strategies await comparable cost analyses that include state-by-state assessment and future projections. Nevertheless, estimated abatement costs for BECCS ($100–400) fall within those of EW. First-of-a-kind DACCS plants ($600–1000 tCO$_2^{-1}$) are 5-fold higher, but those costs will fall as technology and operational experience improves (*43*). CDR strategies such as afforestation ($5–50 tCO$_{2,e}^{-1}$) and soil carbon capture ($0–50 tCO$_{2,e}^{-1}$) have lower U.S. costs than EW (*14*), although, with the yield gain discounts on costs, EW CDR prices could approach $50 tCO$_2^{-1}$ by 2050. Regardless of initial costs, the permanence of soil organic CDR strategies, and their effects on $N_2O$ emissions, remain highly uncertain (e.g., (*44, 45*)). Moreover, unlike these CDR strategies, EW can arguably contribute to greater social equity in decarbonisation benefits, with basalt supply states gaining employment from mineral extraction and processing, and land managers of Corn Belt states gaining carbon removal credits and improved agricultural lands. Employment gains from mineral processing could reverse steady long-term declines in mining of fossil fuels (e.g., quarrying, and coal, oil and gas extraction) due to productivity gains per worker with advances in technology (*46*) and align with a more equitable clean energy future through carbon management (*47*).

The removal of atmospheric $CO_2$ has a net benefit to society (e.g., climate benefits and damage avoided) that can be monetized as the social cost of carbon (SC-$CO_2$) (*48*). Viewing EW CDR as a component of a mitigation portfolio, and taking our lead from regulatory impact analysis, we multiply CDR by EW with a preferred SC-$CO_2$ estimate of $185 per tonne of $CO_2$ to calculate state-level revenue generation potential of EW (**Fig. 5F**). Results indicate EW could provide opportunities for gross revenue generation of more than $50 billion by 2050 (**Fig. 5F**) in states that are perennially important for determining the outcome of U.S. Presidential elections, and control of both houses of congress (Iowa, Michigan, Minnesota, Wisconsin, Pennsylvania). These states also have the lowest initial costs of implementation (**Fig. 5C**) and rural populations facing multiple stressors, including a decline in work, and geopolitical shifts that affect the competitiveness of exports. EW may therefore have an important role of plat in creating create income and jobs within these communities.

This analysis highlights EW as a substantial opportunity for creating a highly competitive multi-billion-dollar CDR industry by ramping up pathways of deployment across agricultural states throughout the Midwest as well as basalt supply states through expansion of the local quarry industry (**Fig. 5F**). Revenue generation in this way allows EW to contribute to market opportunities, employment and rural economic diversification.

**Social license to operate**

Sustainable at scale deployment of EW across the U.S. will impact the environment and must integrate with existing farming practices. This requires societal acceptance from the national- and state political scales through to local communities and individual farmers. Gaining that acceptance requires working with stakeholders and farming communities to identify the conditions under which a social license to operate might exist (49). Proceeding without such a mandate risks significant resistance and conflict (50).

Research on U.S. public perceptions shows that awareness and knowledge of both CDR (51) and EW (52) are very low, hence better information provision and ongoing dialogue with affected stakeholders will be needed. When EW is described in detail to people drawn from a broad cross-section of the public, concerns are raised about possible risks. Concerns include impacts upon the environment, including the oceans and wildlife, as well as a discourse that CDR is only a temporary fix to climate change when the real problem is rising GHG emissions (53). This implies that broad societal support is unlikely to be forthcoming unless EW is developed alongside an ambitious portfolio of conventional climate mitigation measures. Additionally, a well-documented political polarization exists in U.S. attitudes towards climate change (54). Within the agriculture sector, the financial benefit of yield increases, and improved land value, may help align incentives for decarbonizing highly mechanised agriculture using EW.

Efforts to significantly increase mineral production or to establish new quarries may be an issue for some communities across the U.S. Research on the siting of other technologies suggests ongoing and early dialogue with affected stakeholders, equity and community sharing of benefits, monitoring and control of impacts upon local habitats and species, and attention to building trustworthy relationships, are important for generating a social license to operate (55). In principle, EW deployment could bring significant economic and community benefits for both quarrying and agricultural States, but the extent to which decarbonising agriculture will genuinely achieve a just transition is a critical but little understood question (56).

Currently, most farmers in the Corn Belt already apply crushed rocks (limestone) to their fields and soil pH management is viewed as an essential aspect of industrial agriculture and soil health (30). Therefore, multiple key operational aspects of this CDR approach already have buy-in from the most influential stakeholders: farmers and agricultural communities. Rapidly expanding commercial sector activities driving the scaling of EW with basalt and agriculture worldwide, testify to the willingness of farmers, agricultural communities and stakeholders from diverse regions, to engage with this CDR as it matures (57).

**Conclusions**

Transitioning to manage U.S. agriculture using EW with crushed basalt is one of the few practical options for innovation in large-scale farming practices (58) with the potential reverse its contribution to climate change while simultaneously enhancing food, soil and biofuel security. With straightforward technological pathways to upscaling that utilize existing supply chains, EW offers a solution to permanently sequester atmospheric carbon and assist with meeting U.S. net-zero objectives. Despite the large potential deployment area, 70 million hectares of farmland in the Corn Belt alone, with high CDR capacity, EW is currently absent from U.S. government net-zero policy (1-3). However, implementing a stage-gating framework as upscaling proceeds to safeguard against environmental and biodiversity concerns will be essential. Upscaling EW has scope to create a new multi-billion-dollar CDR industry that affords potential opportunities for environmental justice in decarbonisation while also enabling positive action on multiple interacting planetary boundaries. Nonetheless, it is sobering to consider that the required rate of CDR deployment to meet net zero emissions in the U.S., and globally, by 2050 dwarfs the rapid growth of renewable energy and the general pace of our transition away from fossil fuels (2).

**Materials and Methods**

U.S. state-level basalt geochemistry and mineralogy datasets

To determine the mineralogy of U.S. basalts for our EW soil profile simulations, we acquired geochemical data (up to 13 major elements expressed as weight percent oxides) from the literature and two online repositories (59-61) for 9784 mafic rock samples (Error! Reference source not found.**, A**). The samples in our databases were collected by petrologists interested in tectonic history, magmas etc., but we consider that they are representative of the geochemical and mineraological characteristics for potential mined deposits. Total Alkali-Silica (TAS) plots of Na/K oxide vs SiO revealed the majority of classifiable samples (92%) are basaltic, of which 50% are basalts (Error! Reference source not found.**, B**). Geochemical oxide data were converted to "normative" mineralogies following the same procedures as the publicly available spreadsheet (62). These calculations modify those of Johannsen (63) and may differ from simpler algorithms, including Hollocher's procedure using eleven major oxides (64), and from those developed and subsequently modified by Cross *et al.* in the early 1900s (65).

The normative mineralogy approach has the advantage that geochemical data are far more widely available than modal (e.g., X-ray diffraction determined) basalt mineralogies, and this allows us to capture the variability between U.S. states. However, normative mineralogies ("norms") are idealised; for example, secondary minerals, altered/metamorphic minerals, xenoliths, trace carbonates and igneous mineral zoned phenocrysts (e.g., with different internal and external plagioclase) are not accounted for by the normative calculations. As this method is inappropriate for metamorphic rocks, metabasalts are excluded from our study. Additionally, glasses, fast-weathering non-mineral phases present in basalts with chilled margins, are not accounted for. We note that glass has been observed in many of our rocks including the Columbia River basalts; as we do not account for glass our elemental release rates from these rocks are likely be conservative.

The following geochemical data are utilised in Hollocher's spreadsheet (62) and in our Matlab software:

- weight percents of up to 13 major elements expressed as oxides ($SiO_2$, $TiO_2$, $Al_2O_3$, $Fe_2O_3$, $FeO$, $MnO$, $MgO$, $CaO$, $Na_2O$, $K_2O$, $P_2O_5$, $CO_2$ and $SO_3$)
- weight percents of elemental S, F, and Cl if available
- Sr, Ba, Ni, Cr and Zr given in ppm if available.

Missing data are set to zero. Note that reduced and oxidised Fe and S are treated separately. In geochemical analyses, iron is often given as total iron expressed as either $FeO$ or $Fe_2O_3$. Assuming that $FeO$ or $Fe_2O_3$ given alone represent total Fe, we calculated the fraction of reduced Fe as a function of $SiO_2$, $Na_2O$ and $K_2O$ for mafic volcanic rocks (66). Where authors provided both $FeO$ and $Fe_2O_3$, we did not recalculate them. We calculated the chemical index of alteration (67) (CIA) and excluded 17 samples with CIA>40. In the final step prior to calculating the normative mineralogies, we rescaled all our analyses on an anhydrous basis (68) such that the weight percent sum of the oxides above equalled 100%. This helps to ensure consistency of the datasets and follows established protocols for rock classification (68). Following calculation of the norms, 9160 samples with at least 95% normative minerals and <50% quartz were retained.

Boxplots of the oxide data suggest only minor differences between east and west coast mafic rocks (**Error! Reference source not found.**), but western rocks do show a slightly greater range of values for the important plant nutrients K and P (Error! Reference source not found.). There are three K-bearing normative minerals in the scheme used here: orthoclase (K-feldspar), kalsilite and leucite. Of these, normative orthoclase is nonzero for 9140 samples, normative leucite for only one sample, and no samples had normative kalsilite. As orthoclase is a slow-weathering mineral and non-normative phases such as glasses are excluded, our K release is likely to be conservative. Apatite is the only normative P-bearing mineral; it is nonzero for 8752 samples.

For each relatively unaltered mafic rock sample, normative quartz, diopside, wollastonite, hypersthene, orthoclase, ilmenite, olivine, magnetite, and apatite were written to a .csv file along with plagioclase derived from normative albite and anorthite. Because weathering rates increase nonlinearly as the Ca-rich end of the plagioclase solid solution series is approached (69), we classify plagioclase based on the molar Ca/(Ca+Na) percentage ($An_x$) as albite ($An_{0–10}$), oligoclase ($An_{10–30}$), andesine ($An_{30–50}$), labradorite ($An_{50–70}$), bytownite ($An_{70–90}$) or anorthite ($An_{90–100}$). Lower limits are only inclusive for albite, i.e. $An_{50}$ is andesine. Full datasets with sample references and associated Matlab software are available as compressed tar files (Data Files 1-3).

Clustering procedures

The above analysis provided 9784 relative mineralogies for mafic rocks in the continental U.S. (bytownite, labradorite, andesine, diopside, K-feldspar, olivine, apatite, wollastonite, quartz, enstatite, ilmenite, magnetite, albite, oligoclase, anorthite). For each state, we used all the samples falling within its geographical boundary. To avoid computationally expensive simulations for each of the samples (in the 100s), we performed cluster analysis grouping samples with similar mineralogies together. We used two clusters because after performing the analysis for a range of cluster values we observed on a scree plot that the inertia (the sum of squared distances of samples to their closest cluster centre, and a measure of how representative the samples are compared to the cluster they were assigned to) remained constant after the second cluster. By averaging and normalizing the mineral content of the samples in each of two clusters, we obtained two representative mineralogies for each state for 1-D soil weathering model simulations (**fig. S4**). Additionally, during the clustering analysis, mineral proportions were assigned weights according to their weathering potential, for example Kfeldspar was assigned a higher weight than quartz as a more reactive mineral. This ensured that the clusters created better represent the weathering characteristics of the rocks.

Clustering analysis also provided the number of samples $n_i$ that were assigned to each cluster $C_i$, from which we defined the probability of each cluster as $P(C_i) = {n_i}/{\sum_i^2 n_i}$. Hereafter, we used this definition and for each related model output variable V, we calculate its mean value as $V_\mu = \sum_{i=1}^{2} P(C_i) V_i$ and its variance as $V_{\sigma^2} = \sum_{i=1}^{2}(V_i - V_\mu)^2 P(C_i)$, where $V_i$ the variable output obtained by running the model using mineralogy from $C_i$. The mean and variance were then used to calculate variable uncertainties stemming from mineralogy.

State-level rock production.

For each state, we created potential rock production curves *P(y)* for year *y* by tuning the parameters of the logistic function:

$$P(y|m,r,L) = \frac{L}{1 + e^{-r(y-m)}} \qquad \text{Eq. 1}$$

The rationale behind the functional form is that rock production will have a slow roll-out phase of 10–20 years (with delays due to mining licenses, public acceptance, etc), followed by a period of rapid expansion with the opening of new mines, before finally reaching steady-state production rates (**fig. S5**).

Parameter *m* defines the logistic functions inflection point (midpoint) or in this context, the year when the states rate of production maximizes. We hypothesized that states closer to croplands will maximize production sooner while states further away will lag and take over after the initial states encounter diminishing returns. We thus set a base value $m_0$=2041, a scaling constant $m_1$=10 and calculated for each state a normalized (0–1) distance to the croplands *d*. We then calculate the parameter *m* for each state *s* as $m_s = m_0 + m_1 * d(s)$. States close to croplands will maximize rock production at year $m_0$ while states further away at $m_0 + m_1$.

A similar approach was followed for the rate parameter *r* which characterizes the rate of production and was modelled as $r_s = r_0 + r_1 * h(s)$, with $r_0 = 0.22$ and $r_1 = 0.07$. Here *h(s)* is the normalized (0-1) historic production (2010-2018) of crushed traprock for each state (70). The hypothesis is that states that are currently producing large quantities of rock, will be able to ramp up production faster and achieve higher annual rates of rock production.

Finally, the maximum rock production $L_s$ achieved by each state at year 2070 was calculated as a linear combination of historical production (70) and distance to croplands $L_s = 0.7 \times h(s) - 1.7 \times d(s)$. Notice the minus sign, penalizing states away from croplands. Maximum rock production was always normalized so that the sum of production would equal 1 or 2 Gt/y at 2070, depending on the scenario.

Allocation of rock supply states to demand states

Assume *S* source states provide basalt and *P* agricultural targets that can receive it. The source states are represented as single pixels (*n* = 13) of a raster image at the location of the geographic centre of the state, while the targets are all the pixels (n ≈5000) found withing the boundaries of the target states where maize, soy or wheat are present in the CLM land cover representation. Note that locations with more than one crop type, are treated independently as some driving variable are output per crop type from CLM. The optimization was tasked to find the optimal way for the source states to distribute basalt to the agricultural targets which maximizes net carbon sequestration while adhering to certain constrains discussed below. The parameters we are thus trying to obtain values for are the tuples of all possible combinations between source states and targets: $(S_1,T_1),(S_1,T_2),...,(S_1,T_s),(S_2,T_1),(S_2,T_2),...,(S_2,T_s),...,(S_s,T_1),(S_s,T_2),(S_s,T_s)$ , $\forall s \in S, \forall t \in T$.

The tuples can only attain a 1 value if a state does supply basalt to the target, and 0 if not.

In contrast to continuous optimization, discrete optimization (or linear integer programming) is a branch of linear optimization algorithms that deals with problems where the variables being optimized have discrete values. A subcase, applied here, is linear binary optimization where the variables can only obtain 0 or 1 value. We used the PuLP v2.6 Linear Programming API, which utilizes Pythons list comprehensions and concisely creates the thousands of linear constraints required for this problem.

Constraints on rock allocation to croplands

The 1-D soil profile EW model requires a specific basalt is applied to a location at year $y$, continuously until the end of the run. To ensure then that any target field $t$ receives basalt from a single state then for every target $t$ ($\forall t \in T$): $\sum_i^S (S_i, T_t) \leq 1$.

To ensure that the basalt provided by any state $s$ doesn't exceed the states' basalt production $P_{s,y}$ (in t rock) for year $y$, then for every state $s$ ($\forall s \in S$) and year $y$: $\sum_i^t (S_s, T_i) * area(T_i) * appRate \leq P_{s,y}$ where $area$ the area of the target in ha and $appRate$ the basalt application rate set at 40 t/ha.

A third constraint was that the basalt transport distance $D$ is less than 5000 km for every state $s$ ($\forall s \in S$) and every target $t$ ($\forall t \in T$): $(S_s, T_t) * D(S_s, T_t) \leq 5000$.

The function to maximize was the total net CDR (in t CO$_2$) on each year $y$.

$$CDR_{Net}(y) = \sum_{s=1}^{S}\sum_{t=1}^{T}[CDR_{Gross}(S_s,T_t,y) - Tra_E(S_s,T_t,y) - Gri_E(S_s,y) - Spr_E(T_t,y) + Fer_E(T_t,y)] * (S_s,T_t) \quad \text{Eq. 2}$$

Here, $Tra_E$ are the transport CO$_2$ emissions which depend both on source state and target on account of distance travelled, $Gri_E$ the grinding emissions which depended only on the source state assuming grinding takes place at the mining location, the spreading emissions $Spr_E$, depended only on the target where the spreading will take place, and avoided fertilizer emissions $Fer_E$ which depended on the target where the weathering occurs and P/K are released All the processes in tonnes of CO$_2$ and functions of year $y$. For details on each process see respective sections. Fertilizer emissions have the opposite sign to the other secondary processes as the mineral release from EW results in substitution of fertilizer application and thus savings in both emissions and costs.

A final constraint was added on costs. For the optimization we defined total EW costs for year $y$ (in $) as

$$CDR_{Cost}(y) = \sum_{s=1}^{S}\sum_{t=1}^{T}[Tra_C(S_s,T_t,y) + Gri_C(S_s,y) + Spr_C(T_t,y) - Fer_E(y)] \quad \text{Eq. 3}$$
$$* (S_s, T_t)$$

For every state, $s$ ($\forall s \in S$) and every target $t$ ($\forall t \in T$) we restricted costs below certain price points ($C_{thr}$ = \$300/tCO$_2$) as

$$\left[CDR_{Cost}(s,t,y) \big/ CDR_{Net}(s,t,y)\right] * (s,t) \leq C_{thr} \quad \text{Eq. 4}$$

These procedures allow calculation of the average distance from a rock production state to cropland as EW deployment evolves over time.

Feedstock extraction curves for each state and each scenario are given in Error! Reference source not found.. Minnesota and Wisconsin do the heavy lifting with an early ramp-up of production due to their proximity to the corn-belt. Michigan and Virginia follow, with Virginia lagging 5 years behind as it is further from crop areas than Michigan but ramping up production faster as it is already producing 50Mt of crushed rock a year. Washington, Massachusetts, and California will only produce small volumes of basalt to cover mostly in-state requirements, due to their distance to the corn-belt or other major agricultural regions.

Coupled Climate-CN cycle EW Simulations. Our model simulation framework starts with future U.S. climate (2020-2070) from the medium-mitigation future pathway climate (SSP2-4.5) ensemble of CMIP6 runs with the Community Earth System Model v.2 (CESM2) (71). This future climate scenario is at the same time used to drive the Community Land Model v.5 (CLM5), the CESM2 land model, to simulate at high spatial resolution (23 km × 31 km) and temporal (30 min) resolution terrestrial carbon and nitrogen cycling with prognostic crop growth and other ecosystem processes, including heterotrophic respiration (72). CLM5 simulates monthly crop productivity, soil hydrology (precipitation minus evapotranspiration), soil respiration and nitrogen cycling. CLM5 has representation of eight active crop functional types, each with specific ecophysiological, phenological and biogeochemical parameters (72). CLM5 includes CO$_2$ fertilization effects on agricultural systems benchmarked against experiments and observations (73, 74). Atmospheric CO$_2$ increases ~100 ppm from 2015 to 2070, as defined by SSP2-4.5. In our CLM5 simulations with rising CO$_2$ and climate change drive increases in crop NPP and reductions in evapotranspiration, which can facilitate weathering in our soil profile EW model (16). We initialized CLM5 simulations in 2015 using fully spun-up conditions from global runs at ~100 km × 100 km resolution, adding an extra 200-year spin-up in the regional set-up to stabilize the CN pools to the higher resolution setting.

CLM5 includes an interactive nitrogen fertilization scheme that simulates fertilization by adding nitrogen directly to the soil mineral nitrogen pool to meet crop nitrogen demands using both synthetic fertilizer and manure application (72). We kept land use and land cover constant to 2015 to be able to spatially track basalt application through the years. Synthetic fertilizer application is prescribed by crop type based on the Land Use Model Intercomparison Project (75), with N-fertilizer rates increasing by 3% per decade from 2020 to 2050 in agreement with current N-fertilizer usage increases (76, 77), and then stabilize from 2050 to 2070. Average U.S. CLM5 fertilizer application rates in 2020 (102 kg N ha$^{-1}$yr$^{-1}$) are consistent with current practices (76). Organic fertilizer is applied at a fixed rate (20 kg N ha$^{-1}$yr$^{-1}$) throughout the simulations.

CLM5 tracks nitrogen content in soil, plant, and organic matter as an array of separate nitrogen pools and biogeochemical transformations, with exchange fluxes of nitrogen between these pools (72). The model represents inorganic N transformations based on the DayCent model, which includes separate dissolved $NH_4^+$ and $NO_3^-$ pools, as well as environmentally controlled nitrification, denitrification and volatilization rates (38). To model the effect of basalt addition on fluxes of $N_2O$, NO and $NH_3$ from soil, we included the updated denitrification DayCent module (33) with some other implementations (34). Our parameterization only considers the possible effect of increased soil pH from basalt application on direct agriculture emissions of $N_2O$, NO and $NH_3$ volatilization. Indirect soil $N_2O$ emissions are not explicitly modelled, but the small bias is likely marginal, given that these emissions account for less than 5% of total agricultural $N_2O$ emissions (78, 79). Cropland CLM5 soil nitrogen emissions are within the range of estimates in U.S. based on field observations, bottom-up inventories and other land surface models (34).

Air Quality Simulations. To assess the effect of changes in soil nitrogen emissions from EW with croplands on air quality, we used CESM2.2. For the atmospheric component, we employed CAM-chem version 6 with the MOZART Troposphere Stratosphere (TS1) chemistry mechanism (80), which includes 221 gas phase and aerosol species and 528 chemical and photochemical reactions. Aerosol concentrations and size distribution are derived from the four-mode Modal Aerosol Model (MAM4) (81). We included the Model for Simulating Aerosol Interactions and Chemistry (MOSAIC) with MAM4 to simulate nitrate aerosol (82). CAM6-chem is coupled to the interactive CLM5, which provides biogenic emissions and handles dry deposition.

We performed two sets of experiments: Control (present-day climate and future emissions with no basalt application) and EW (present-day climate and future emissions with basalt application). For future emissions, we used anthropogenic and biomass burning emissions from the inventory developed for CMIP6 SSP2-4.5 (83). Soil agriculture NO and $NH_3$ emissions were obtained from our CLM5 EW 2 Gt yr$^{-1}$ modelling experiment, and regridded to match our CAM6-chem model resolution by using bilinear interpolation; CMIP6 anthropogenic NO and $NH_3$ emissions were adjusted for not doubling counting agriculture emissions. Table S1 summarizes the main anthropogenic emissions for short-lived air pollutants projected over the United States, and the soil agriculture NO and $NH_3$ emissions. For climate, we run CAM6-chem nudged to the Modern-Era Retrospective analysis for Research and Applications (MERRA2) meteorological fields from 2015 to 2018 (84) to reduce variability related to dynamic simulations in the atmospheric model (CAM6) and fully focus on surface $O_3$ and $PM_{2.5}$ air quality changes from EW.

We performed 4-year simulations, with the first year as model spin-up, for 2030, 2050 and 2070 emission conditions. Simulations were performed at the horizontal resolution of 0.9° x 1.25° horizontal resolution (about 100 km × 100 km) and vertical resolution of 32 vertical layers from the surface to the pressure height of 3.6 hPa (~45 km), with a time step of 30 min. CESM2 simulations have been extensively evaluated by comparison with satellite, sonde, aircraft and ground observations on regional and global scales (e.g., 38, 80, 85, 86).

Air pollution effect on crop yields. To study the effect of projected changes in surface $O_3$ levels and crop yields from EW, we used the ozone exposure metric AOT40 established to protect crop health:

$$\text{AOT40} = \sum_{i=1}^{n} \left([O_3]_i - 0.04\right) \qquad \text{Eq. 5}$$

where $[O_3]_i$ is the hourly mean ozone mixing ratio in ppm during the twelve hours of local daylight (08:00-19:59); $n$ is the number of hours in the 3-month growing season defined as the three months prior to the start of the harvest period for each crop. We considered June–August for maize, soybean and wheat in the USA. AOT40 is given in units of ppmh.

The relative yield (RY) for each crop was then calculated using AOT40 and represents the reduction in yield due to $O_3$ exposure relative to a case where AOT40 is zero.

$$RY_{maize} = 1 - 0.0124 \text{AOT40} \quad \text{(ref (36))} \qquad \text{Eq. 6}$$

$$RY_{wheat} = 1 - 0.0163 \text{AOT40} \quad \text{(ref (87))} \qquad \text{Eq. 7}$$

$$RY_{soy} = 1 - 0.0090 \text{AOT40} \quad \text{(ref (36))} \qquad \text{Eq. 8}$$

Economic crop gains. To estimate economic yield gains from changes in crop yields, we used the US Department of Agriculture National Agricultural Statistics Services per state dataset (88). We extracted crop yield (BU/acre), price ($/BU) and area (acre) for maize, wheat and soybean reported in 2020 for the main states in the analysis. Crop prices were adjusted to consider market changes in 2030, 2050 and 2070 using the FAO market producer price indexes for a Towards Sustainability scenario in the U.S. (89).

To estimate the economic gains from increases in crop yield from basalt applications (**Fig. 5**), we used maize and soybean relative yields obtained from EW field trials at the Energy Farm in the heart of the Corn Belt (13), i.e., 12% (8−16) for maize and 16% (12−20%) for soybean. Field data for wheat were not available. To estimate the economic gains from avoided surface $O_3$ pollution (**Fig. 4**), we used the relative yield changes estimated from our modelled AOT40 data.

Soil Profile EW Modelling.

Our analysis uses a 1-D vertical reactive transport model for rock weathering with steady-state flow and transport through a series of soil layers (6). The transport equations include source terms representing rock grain dissolution and alkalinity within the soil profile. Contributions to alkalinity come from crop productivity (acidifier) and biogeochemical transformations of nitrogen fertilizers.

The core model accounts for changing dissolution rates with soil depth and time as grains dissolve, and chemical inhibition of dissolution as pore fluids approach equilibrium with respect to the reacting basaltic mineral phases, and the formation and dissolution of pedogenic calcium carbonate mineral in equilibrium with pore fluids (6). Simulations consider for each state incorporate basalts with representative specified mineralogies, as determined by cluster analyses of our normative mineralogy datasets (Error! Reference source not found.).

We model EW of a defined particle size distribution (psd) with the theory developed previously (6). As the existing psds at each soil layer are at different stages of weathering, the combined psd at each level, and for each mineral, is calculated and tracked over time. We account for repeated basalt applications by combining the existing psd with the psd of the new application. Simulated mineral dissolution fluxes from the model output were used to calculate the release of P and K over time. Mass transfer of P within the relatively more rapidly dissolving (90) accessory mineral apatite is calculated based on the P content of the rock and the volume of bulk minerals dissolved during each time step.

The mathematical model combines a multi-species geochemical transport model with a mineral mass balance and rate equations for the chemical dissolution of basaltic mineral phases. The model includes an alkalinity mass balance that includes the effect of fertilizer applications and soil N cycling and dynamic calculation of pH in soil pore waters. The main governing equations are the following.

Transport equation. The calculated state variable in the transport equation is the dissolved molar equivalents of elements released by stoichiometric dissolution of mineral $i$, in units of mole L$^{-1}$. $\phi$ is volumetric water content, $C_i$ is dissolved concentration (mole L$^{-1}$) of mineral $i$ transferred to solution, $t$ is time (mths), $q$ is vertical water flux (m y$^{-1}$), $z$ is distance along vertical flow path (m), $R_i$ is the weathering rate of basalt mineral i (mole per litre of bulk soil mth$^{-1}$) and $C_{eq_i}$ is the solution concentration of weathering product at equilibrium with the mineral phase $i$ (Eq. 9). Values for $C_{eq}$ for each of the mineral phases in the basalt grains are computed by calibrating the results of the performance model against those of a 1-D reactive transport model, with detailed geochemical speciation and secondary mineral equilibria, as described previously (6). This calibration is achieved by minimizing a cost function based on the differences in the outputs of the two models over the period of 1 year. These values of $C_{eq}$ are computed for a range of temperatures and flowrates: specifically, temperature (10, 15, 20, 25°C) and flowrate (0.25, 0.5, 1, 1.5, 2, 2.5, 3 m/yr) for a profile consisting of particles of radius 10μm (**figs. S6, S7**). This particle size provides a benchmark for the maximum dissolution rate, which is lower than that for the larger particle sizes considered in the simulations. Therefore, this representation of weathering kinetics provides a conservative lower bound on dissolution rate due to its inhibition upon approaching solubility equilibrium, represented by $C_{eq}$.

Rates of basalt grain weathering define the source term for weathering products and are calculated, for each component mineral and its relative reactive area in the basalt grains, as a function of soil pH, soil temperature, soil hydrology, soil respiration and crop net primary productivity (NPP) (6). The vertical water flux is zero when pore water content is below a critical threshold for vertical flow. Weathering occurs under no-flow conditions and the accumulated solutes in pore water are then advected when water flow is initiated under sufficient wetting, tracked using a single bucket model.

$$\phi \frac{\partial C_i}{\partial t} = -q \frac{\partial C_i}{\partial z} + R(pH, tmp)\left(1 - \frac{C_i}{C_{eq}(tmp, q)}\right) \quad \text{Eq. 9}$$

Mineral mass balance. The change in mass of basalt mineral $i$, $B_i$, is defined by the rate of stoichiometric mass transfer of mineral $i$ elements to solution. Eq. 10 is required because we are considering a finite mass of weathering rock, which over time can react to completion, either when solubility equilibrium between minerals and pore water composition is reached, or when applied basalt is fully depleted.

$$\frac{\partial B_i}{\partial t} = -R(pH, tmp)\left(1 - \frac{C_i}{C_{eq}(tmp, q)}\right) \quad \text{Eq. 10}$$

Removal of weathering products. The total mass balance over time (Eq. 11) for basalt mineral weathering allows calculation of the products transported from the soil profile. The total mass of weathering basalt is defined as follows where m is the total number of weathering minerals in the rock, $t_f$ is the duration of weathering and $L$ is the total depth of the soil profile (m). We define $q$ as the net monthly sum of water gained through precipitation and irrigation, minus evapotranspiration, as calculated by CLM5.

$$\text{Total weathered Basalt} = \sum_{i=1}^{m} [\phi \int_{z=0}^{L} C_i(t,z) \, dz + q \int_{t=0}^{t_f} C_i(t,L) \, dt] \quad \text{Eq. 11}$$

Modelling Soil Nitrogen Effects on EW. Simulation of the effect of nitrogen cycling processes on EW is via sixteen stoichiometric nitrogen transformations included with the CLM5 code that influence the soil weathering environment (16). The modelling accounts for 20 layers in the soil profile at each location with a monthly time-step; variables passed from CLM5 by time and depth to the 1-D EW model are given in ref (16). At each depth, we compute nitrogen transformation effects on soil water alkalinity with reaction stoichiometries that add or remove alkalinity. Together with soil $CO_2$ levels, this affects pore water pH and the aqueous speciation that determines mineral weathering rates. This allows us to account mechanistically for the impact of soil acidification, including that from N fertilization on EW and CDR, a potential source of nitric acid weathering at low pH that limits CDR by EW in cropland. Dynamic modelling at monthly time-steps resolves seasonal cycles of CDR via alkalinity fluxes and soil carbonate formation/dissolution in response to future changes in atmospheric $CO_2$, climate, land surface hydrology, and crop and soil processes. The effect of the nitrogen cycle on the soil acidity balance is derived from nitrogen transformations associated with the production or consumption of hydrogen ions.

We assigned a stoichiometric acidity flux $\Delta H_{i,N}$ (mol H$^+$ mol$^{-1}$ N) to each nitrogen flux $F_{i,N}$ (gN m$^{-3}$soil s$^{-1}$) calculated by the CLM5 code. The product ($F_{i,N} \cdot \Delta H_{i,N}$), with appropriate unit conversions, gives the acidity flux during the time-step $\Delta t$ (month) for the $i^{th}$ reaction of the CLM5 nitrogen cycle. Their sum (Eq. 12) is, therefore, the total change in acidity $\Delta \text{Acidity}_N$ due to the CLM5 nitrogen cycle:

$$\Delta \text{Acidity}_N = \sum ( F_{i,N} \, \Delta H_{i,N}) / 14.0067 \, \Delta t \quad \text{Eq. 12}$$

where 14.0067 gN mol$^{-1}$ N is the atomic weight of nitrogen and the time-step is one month. Along with the Ca, Mg, K, and Na ions released from weathering the applied minerals, $\Delta \text{Acidity}_N$ contributes a negative term to the soil water alkalinity balance used to calculate the soil pH (Eq. 13):

$$\text{Alk}_t = \text{Alk}_{t-1} + 2 \cdot (\text{Ca}_{weath} + \text{Mg}_{weath}) + \text{K}_{weath} + \text{Na}_{weath} - \Delta \text{Acidity}_N \quad \text{Eq. 13}$$

This pH value is one variable that is accounted for in the rate laws for mineral dissolution and therefore influences the rate of basalt weathering and the resulting net alkalinity that is produced at each depth within the soil profile, and that contributes to CDR (6, 16).

Modelling Soil Profile pH and its Effects on EW. The initial alkalinity profile in each grid cell is determined from the starting soil pH and the $PCO_2$ profile at steady-state based on spin-up of the model with average biomass production and soil organic matter decomposition that reflects the long-term land use history of a particular location. Alkalinity mass and flux balance for an adaptive time-step accounts for alkalinity and acidity inputs from 1) mineral dissolution rates and secondary mineral precipitation (pedogenic carbonate), 2) biomass production and decomposition (91) and 3) biogeochemical N transformations. The soil pH profile is determined from an empirical soil pH buffering capacity (92) relating soil pH to the alkalinity at each depth. The soil $PCO_2$ depth profile of a grid cell is generated with the standard gas diffusion equation (93), scaled by monthly soil respiration from CLM5.

The vertical root respiration profile based on the monthly total value is assumed to be distributed proportionally to the root biomass. A traditional approach is taken to estimate the vertical root distribution by applying an exponential root density profile given a column total (94). The root density profile is then given by Eq. 14:

$$\rho_{rd}(t,z) = A(t)e^{-az} \qquad \text{Eq. 14}$$

where $z$ is the depth; $a = 1/L$, where L is a vegetation dependent e-folding length scale; and *A(t)* is the surface root density. The rooting depth (*d*) is typically defined as the depth that contains 95% of the root biomass, which can be shown to be equal to (Eq. 15):

$$d = \frac{-\ln(1-0.95)}{a} = \frac{3}{a} \qquad \text{Eq. 15}$$

Given that the root respiration ($\rho_{rr}$) is proportional to the root density and $A_{rr}$ is the surface respiration, then

$$\rho_{rr}(t,z) = A_{rr}(t)e^{-3z/d} \qquad \text{Eq. 16}$$

Integrating Eq. 16 over the soil depth gives the total root respiration $RR(t) = A_{rr}d/3$. And therefore

$$\rho_{rr}(t,z) = \frac{3RR(t)e^{-3z/d}}{d} \qquad \text{Eq. 17}$$

The above source term (Eq. 17) is used in a transport equation (Eq. 18) to track the alkalinity contribution from the root respiration/acidifier.

$$\phi\frac{\delta A}{\delta t} = -q\frac{\delta A}{\delta z} + \frac{3RR(t)e^{-3z/d}}{d} \qquad \text{Eq. 18}$$

At any particular location, the soil solution is in dynamic equilibrium with dissolved inorganic carbon species and the values of gas phase soil and atmospheric $PCO_2$. The relative change induced by weathering will be the consumption of $H^+$ and the production of $HCO_3^-$.

Using this modelling framework, we analyzed a baseline application rate of 40 t ha$^{-1}$ yr$^{-1}$ (equivalent to a <2 mm layer of rock powder distributed on croplands) to U.S. corn-belt croplands. Similar road/rail transport of mass occurs in reverse during transportation of yield from field to market, e.g., with sugarcane yields of up to 91 t ha$^{-1}$ in Florida (95) and perennial bioenergy crop *Miscanthus*, relevant to U.S. biofuel implementation programmes (~10-30 t ha$^{-1}$) (96).

Costs and Carbon emissions of EW Operations. Spreading Costs: We set a value of $5.5 per ton of rock as the basis for spreading costs, based on the 2020 values for lime spreading in the United States. In that year, we assumed that 40% of the cost was for wages, 30% for diesel fuel, and 30% for capital costs. To project spreading costs from 2020 to 2070, we used the E3-US model outputs to obtain state-specific time series of median salary, diesel, and electricity prices. These values were normalized for 2020 and used to calculate state-specific spreading cost projections based on the cost breakdown. We assumed a modest decarbonization of the operations in the agricultural sector, with electricity replacing diesel. The transition starts at 0% in 2020 and reaches 38% and 100% by 2050 and 2070, respectively.

Spreading Emissions: Following (97), we set spreading energy requirements at 21 kWh/t rock and emissions for a diesel operated spreader at 238 g$CO_2$/kWh; when combined we acquire spreading emissions at 0.005 t $CO_2$/t rock. Given the crop area and application rate, we can calculate the total rock requirements and, consequently, the total spreading emissions. With the implementation of decarbonization, we transition from diesel to electric machinery. The emissions from electric-power machinery are derived from the life cycle emissions of each state's electricity grid.

Quarrying and Grinding Costs and Energy

Energy and financial costs associated with the extraction of rock were derived from ref (98), for which we assume the energy costs associated with the detailed UK case-study are consistent with those in the US. The framework for technoeconomic assessment was adapted for enhanced rock weathering from ref (99).

Operating labour and supervision were calculated using working hours from ref (98) and US average salaries based on roles (100). Administration was assumed 5% of operating labour. Fixed operating costs associated with plant, buildings or equipment were converted from £$_{2010}$ to $$_{2020}$ values using an average 2010 exchange rate of 1.546$ to £, a capital location factor for Western Europe of 1.06, and the Chemical Engineering Plant Cost Index (551 in 2010, and 596.2 in 2020) (101), assuming that these costs are typically a function of capital expenditure. 2020 US average diesel and industrial electricity unit costs were used (102). Average US electricity costs were used. Total maintenance costs were assumed to be 2% of total plant costs (**Error! Reference source not found.**).

Capital costs for crushing were converted from £$_{2010}$ to $$_{2020}$ values using an average 2010 exchange rate of 1.546$ to £, a capital location factor for Western Europe of 1.06, and the Chemical Engineering Plant Cost Index (551 in 2010, and 596.2 in 2020). While missing from ref (98), we assumed an Engineering, procurement, and contractors cost of 8% and a contingency of 5% (given that rock extraction is a highly mature and globally scaled industry). Additional owners costs of 20% total plant costs (not included in ref (98)) were also incorporated.

Grinding of rock was not considered in ref (98), but have been incorporated by using free on board costs of grinding, which were calculated by scaling the cost of reference equipment (a semi-autogenous mill: reference cost of $1.7M for 300 kW and scaling exponent of 0.31, and a motor: reference cost of $260k for 2,225 kW and a scaling exponent of 0.81, both to a maximum of 4,500 kW with a labour and maintenance factor of 2.00 (103). Bare erected costs were calculated using the Chemical Engineering Plant Cost Index (1000 for the reference, and 596.2 in 2020). For total power requirements greater than the maximum of a single unit, multiple fully scaled units were used. We assumed an Engineering, procurement, and contractors cost of 8% and a contingency of 5% (given that rock extraction is a highly mature and globally scaled industry). Additional owners' costs of 20% total plant costs were also incorporated (Table S3).

The levelized cost of production (LCOP, Eq. 19) was calculated using the method detailed in ref (104), assuming a capital charge factor (CCF) of 0.11. Capacity factors (CF) were assumed to already be accounted for in production data. a discount rate of 12%, an inflation rate of 2%, an escalation rate of 1%, and a capital service life of 25 years was assumed to calculate a levelization factor (LF, ~1.24). A summary of the costs of production are included in **Error! Reference source not found.**.

$$LCOP = \left(\frac{(CCF \cdot TOC) + (OPEX_{fixed}) + (OPEX_{var} \cdot CF)}{P \cdot CF}\right) \cdot LF \qquad \text{Eq. 19}$$

As mentioned, all costs (labour, electricity, diesel) reflect $\$_{2020}$, averaged over the United States. To obtain state-specific costs over the 2020-2070 period, we averaged and normalized E3-US price projections for each state for the year 2020, and scaled accordingly.

Grinding energy requirements E in kWh/t were obtained from ref. (105) as:

$$E = 4.76 \times M \times \left(\frac{1}{x_{prod}^{f(x)}} - \frac{1}{x_{feed}^{f(x)}}\right) \qquad \text{Eq. 20}$$

where $M$ is the work index (assumed 19.4 kWh/t), $x$ is the P80 in µm of the feed or product (in this study 100 µm), and $f(x) = -0.295 + \left(\frac{x}{10^6}\right)$.

Transportation Costs and $CO_2$ Emissions

We analyzed transportation costs and $CO_2$ emissions for EW deployment in the U.S. by following a similar approach as in our previous studies (6, 16, 23). We estimated costs and emissions for basalt transport by assimilating projected data of fuel/electricity prices, wages, as well as the electricity production mix and the corresponding emission factors at a state level. This data was obtained from the E3-US macro-economic model that considers 1.5°C energy policies scenarios (106).

In terms of fuel efficiency, for freight road transport, we used 2.1 kWh/mile (1.3 kWh/km) for heavy electric vehicles (107, 108). For heavy diesel fuel trucks in 2020, we used 5.8 miles per US gallon (2.51 km/litre) (109). For freight rail fuel efficiency in the United States we used data from historical records at 472 tonne-miles per gallon (110, 111).

To estimate transport costs per tonne-kilometer (t-km), we considered standard road and rail cost models, including wages, fuel, maintenance, and depreciation (112, 113). We assumed that electric HGV will be readily- available from 2030, achieving a 90% market share by 2050 (106, 108, 114). We modelled the rail network decarbonization transition by adding a 20-year lag to the electric HGV scenario (111). We then applied a projected shared pathway of both transport types to estimate transport costs (**fig. S8**).

To calculate transport distances we used geographical data to identify available sources of basalt rock (115) and cropland areas and performed transport network analysis with least cost path algorithms from ArcGIS model builder (116), which looks for and measures the smallest, average and longest transport routes between multiple origins and destination sources. The transport analysis returned transport cost/emissions per t-km between each rock state and all crop sites. This distance tuples were subsequently used in the optimization algorithm.

Finally, we utilized a combination of truck and train modes for transporting basalt to the crops, depending on the distance from the rock source. For distance bands ranging from 50 km to 1750 km, the proportion of truck usage varied from 0.96 to 0.31, respectively (117). This reflects that rail is more cost effective than trucks over longer distances.

The average initial cost of transport for 2020 ranged from $0.07 to $0.028 per t-km for road/rail transport, while average road carbon emissions for all the states in 2020 were around 50g $CO_2$ per t-km, and 15g $CO_2$ per t-km for rail transport. The cost and carbon emissions of both transport modes will drop significantly in the coming decades, with the introduction of electric HGV but a much slower rollout of the electrification of the rail network in the U.S. (109).

Modelling EW Effects on Riverine Carbonate Chemistry.

Our model for estimating carbonate saturation state in the rivers uses well-documented aqueous geochemical datasets in the contiguous U.S., where United States Geological Survey (USGS) has been monitoring the solute composition and discharge rates of the US streams with a high temporal resolution and a high spatial coverage for many decades (118). From USGS, we collected a suite of aqueous species and parameters, including alkalinity, pH, calcium, water temperature, salinity, and discharge rate. Sites with multiple measurements in a day are averaged daily for each parameter, after which all parameters are merged together based on the unique site number and sampling date and any rows containing NA values are deleted. After data filtering, the average monthly value of each parameter for each river is calculated by aggregating the samples by month using a discharge-weighted approach. Also note that for pH values, we first converted original pH values from $-\log_{10}([H^+])$ scale to the $[H^+]$ concentration scale (mol/L) before data aggregation and then converted them back to $-\log_{10}([H^+])$ in the final step for calculating river omega values.

We then delineate the corresponding watershed for each river site using GRASS GIS(119) based on the stream flow direction at a 30" resolution from HydroSHEDS (120). Specifically, we use the GRASS module r.accumulate to derive the stream network and flow accumulation over the northern America from stream direction. Next, we snapped each river gauge to the newly calculated stream network. We then use the module r.water.outlet to create the watershed basin for each snapped stream gauge. The watersheds that are completely located within the states where EW is implemented are selected, which are used to calculate the time-series water infiltration flux (from climate model data) that drains into each river site. Finally, we have a dataset containing 863 river sites (Fig. 2A) with complete hydrogeochemical parameters and watershed boundaries that will be fed into the carbon saturation state (CSS) model. Using the HydroSHEDS stream network, we also delineate 6 large watersheds (including Mississippi, Colorado, Columbia, Sacramento, Lawrence, and Nelson) in North America (Fig. 2B). We then calculate the fluxes of bicarbonate, Mg, Ca, Na, K, Fe, and Si (derived from EW modeling) for each big watershed through time (Error! Reference source not found.**, S10**). These fluxes will be fed into the CSS model to update the carbonate saturation states ($\Omega$) of each river site through time.

Initial and time-series $\Omega$ values during EW are calculated based on solute chemistry, temperature, and salinity according to Eq. 21:

$$W = \frac{[Ca^{2+}][CO_3^{2-}]}{K_{sp}} \qquad \text{Eq. 21}$$

where $K_{sp}$ represents the apparent solubility product for calcite corrected for site-specific temperature and salinity(121) and brackets denote concentration.

We first calculate dissolved inorganic carbon (DIC), $[CO_3^{2-}]$ and $K_{sp}$ using the seacarb package in R (122), considering the impact of both temperature and salinity. Combining $[CO_3^{2-}]$, $K_{sp}$ and $[Ca^{2+}]$, we calculate the initial $\Omega$ for each site following equation S1. The distribution of $\Omega$ in rivers/streams is right-skewed, and the majority of the $\Omega$ values are smaller than 5 (inset in Fig. 2A). To model the evolution of monthly $\Omega$ values in each river site through time, we first used linear interpolation to build the relationship between the monthly background concentration of aqueous species (including Ca, DIC, alkalinity, and salinity) and river discharge for each single watershed. Then we used this relationship to calculate the monthly concentration of these aqueous species in each river site scaled by the water infiltration rate (derived from the land model) through time. Note that these concentration values are treated as the new background values in the river before receiving the EW products. Afterwards, we calculate the extra concentrations of Ca, alkalinity, and salinity generated by EW at each watershed for each month through time and added those extra concentrations to the corresponding background aqueous species concentration to obtain the updated river chemistry.

Finally, we calculate the new $\Omega$ values using DIC, alkalinity, Ca concentration, salinity, and water temperature through time following the method of ref (25). For both the 1 Gt yr$^{-1}$ and 2 Gt yr$^{-1}$ rock extraction scenarios, most river sites experience $\Omega$ values smaller than 5 from 2020 to 2070 (Error! Reference source not found.). This highlights that the transport of dissolved constituents in surface waters is unlikely to be a primary bottleneck limiting the CDR potential of EW.

Estimating the impact of ocean carbon storage on CDR efficiency. We evaluate the CDR efficiency by EW on the U.S. croplands using a 'carbon-centric' version of the Grid Enabled Integrated Earth system model — cGENIE. The ocean physics and climate model components of cGENIE comprise a reduced physics (frictional geostrophic) 3-D ocean circulation model coupled to a 2-D energy-moisture balance model and a dynamic-thermodynamic sea ice model (123). Heat, salinity, and biogeochemical tracers are transported via parameterized isoneutral diffusion and eddy-induced advection (124). The ocean model exchanges heat and moisture with the atmosphere, sea ice, and land while being forced at the ocean surface by zonal and meridional wind stress according to a specified static wind field. Heat and moisture are horizontally mixed throughout the atmosphere and exchange heat and moisture with the ocean and land surfaces, with precipitation occurring above a given relative humidity threshold. The sea ice model tracks horizontal ice transport and exchanges of heat and fresh water, using the thickness, areal fraction, and concentration of ice as prognostic variables. Full descriptions of the climate model and ocean physics can be found in (123, 124). The ocean model is configured here as a 36 × 36 equal-area grid (uniform in longitude and sine of latitude) with 16 logarithmically spaced depth levels and seasonal forcing at the ocean surface.

The ocean and sediment biogeochemistry modules in cGENIE control air-sea gas exchange, the transformation and repartitioning of biogeochemical tracers within the ocean, and the impacts of shallow sediment diagenesis on calcium carbonate formation/dissolution and burial. The ocean biological carbon pump is driven by a parameterized uptake rate of nutrients in the surface ocean, with this flux converted stoichiometrically to biomass that is then partitioned into particulate or dissolved organic matter for downstream advective transport, sinking, and remineralization within the ocean interior. Dissolved organic matter is transported with the ocean circulation and decays according to a specified time constant, while particulate organic matter is instantaneously exported from the surface ocean and is remineralized within the ocean interior following an exponential decay function with a specified remineralization length scale. The ocean biogeochemistry also contains a fully coupled carbonate system, which tracks individual dissolved inorganic carbon (DIC) species, dissolved alkalinity, and ocean pH. Calcium carbonate forms in surface ocean grid cells at a stoichiometric ratio with organic matter production (the so-called "rain ratio") and is exported as a solid species and is dissolved in the ocean interior or shallow marine sediments depending on ambient temperature, pressure, and carbonate chemistry (125, 126). A simple scheme for shallow sediment diagenesis allows us to run the ocean alkalinity cycle as an open system, with delivery from weathering of the land surface and ultimate burial as calcite ($CaCO_3$) in marine sediments). More detailed description and validation of the ocean and sediment biogeochemistry in cGENIE is provided in (127, 128). We implement a simple model of carbon exchange with the terrestrial biosphere in which aboveground biomass (vegetation) and soil carbon are treated as global pools that respond to temperature and atmospheric $pCO_2$ (e.g., a "slab" or "box" terrestrial biosphere) (27).

The model climate system and ocean carbonate/alkalinity cycle are spun up to steady state using a two-stage procedure. First, the model is run as a closed system for 20 kyr with atmospheric abundances of $CO_2$, $CH_4$, and $N_2O$ imposed at preindustrial values to bring the ocean-atmosphere system and shallow sediments into steady state. This run is used to diagnose the approximate steady state burial flux of calcium carbonate in marine sediments, which is then imposed as a weathering flux of calcium and alkalinity in a second stage spin up in which the ocean and sediments are allowed to evolve as an open system. The second stage spin up is run for 20 kyr.

All subsequent simulations are branched from the open system spin up at model year 1765 and run to year 2100 according to the SSP2-45 scenario for atmospheric $CO_2$, $CH_4$, and $N_2O$ (129). Time-varying atmospheric abundances of $CH_4$ and $N_2O$ are imposed according to the SSP2-45 trajectory for all simulations, while atmospheric $CO_2$ abundance is emission-driven. The emission trajectory for SSP2-45 is first computed by the model by prescribing the atmospheric $CO_2$ trajectory for that scenario, with all subsequent runs utilizing the emission trajectory diagnosed in cGENIE for the SSP2-45 pathway.

Our simulations of carbon dioxide capture through EW on croplands in the U.S. reflect riverine delivery of alkalinity (here assumed to be the sum of Mg and Ca) and DIC from 6 watersheds from U. S. riverine data. The riverine alkalinity (**Error! Reference source not found.**) and DIC flux data are averaged every 5 years and fed to 8 consecutive cGENIE runs from 2030 to 2070 as the forcings of DIC and alkalinity fluxes at the surface ocean grid cells corresponding to 6 river mouths (Error! Reference source not found.) in addition to the $CO_2$ removal flux from the atmosphere which corresponds to the $HCO_3$ fluxes (**figs. S9, S10**).

We also simulate a hypothetical carbon cycle intervention meant to represent direct CDR or mitigation of emissions more than that implied by SSP2-45 pathway. For this, we reduce $CO_2$ emission rates by a specified value corresponding to EW in the U.S. croplands except without any addition of alkalinity and DIC to the ocean, relative to the control emission rates for SSP2-45. This simulation is referred to as "baseline" or "base" CDR experiment. By comparing $CO_2$ leakage between EW and baseline experiments we can evaluate the leakage specifically caused by addition of DIC and alkalinity to the ocean through EW (Fig. 3).

U.S. Energy and economic forecasts

Projections of costs are based on a global policy scenario consistent with limiting global warming above pre-industrial level to 1.5°C (without overshoot). The scenario is an updated version of that used in ref (17), using the non-optimization integrated assessment model framework E3ME-FTT-GENIE (18), covering global macroeconomic dynamics (E3ME), S-shaped energy technological change dynamics (FTT), and carbon cycle and climate system (GENIE). The scenario incorporates a range of market-based and regulatory policies, including carbon pricing, energy efficiency standards and incentives for low- and zero-carbon technology uptake in the power sector, for personal transport and buildings. It also includes an estimate of the short-term impacts of the Covid-19 pandemic on the economy and energy system. Most of the climate-related policies employed are already implemented in at least one jurisdiction, and here we assume they are implemented in all nations (which is necessary to meet the temperature target). The resulting modelled global greenhouse gas emissions reach net-zero shortly after 2050, driven by electrification of secondary energy use and near-complete decarbonisation of the power sector, which also features some bioenergy with carbon capture and storage to compensate remaining emissions, mostly from agriculture and industrial processes. Bioenergy use is also increased in aviation and freight transport, but global (modern plus traditional) bioenergy consumption does not exceed 150 EJ in any year.

The scenario is disaggregated to state resolution for the USA using data and projections from the E3-US model (130). E3-US follows the approach used in the global E3ME model, including high disaggregation into 70 econometrically specified sectors defined in each state. E3-US also incorporates an integrated treatment of the energy system, including a bottom-up representation of the power sector. The model was designed to address questions relating to fiscal policy and redistribution across states, including carbon taxes and other environmental fiscal reform and the impacts of new energy regulation, energy efficiency measures, feed-in-tariffs and support for new technologies.

Prices at state level are estimated by mapping the national-level figures for the U.S. to the state-level data, while maintaining overall consistency with national outcomes at sectoral level.  For example, a RAS procedure ensures that the volume of electricity generated by each technology in each of the US's three grids sums to the national total, while still meeting the expected levels of electricity demand. Electricity prices in each state are then determined by taking a weighted average of the levelized costs of each generation technology used in the relevant grid, adjusting for state taxes.

**ACKNOWLEDGMENTS**

We thank Unnada Chewpreecha for assistance with the U.S. energy and economic modelling analysis.  We acknowledge the World Climate Research Programme's Working Group on Coupled Modelling responsible for CMIP and thank the climate modelling groups for producing and making available their model output. For CMIP the US Department of Energy's Program for Climate Model Diagnosis and Intercomparison provides coordinating support and led development of software infrastructure in partnership with the Global Organization for Earth System Science Portals

**Funding**: Supported by the Leverhulme Trust, Leverhulme Research Centre grant RC-2015-029. UKRI funding under the UK Greenhouse Gas Removal Programme (BB/V011359/1, DJB; NE/P019943/1, NE/P019730/1, P.R.). M.V.M. acknowledges funding from the UKRI Future Leaders Fellowship Programme (MR/T019867/1) and high-performance computing support from Cheyenne (doi:10.5065/D6RX99HX) provided by NCAR's Computational and Information Systems Laboratory, sponsored by the National Science Foundation. NJP acknowledges support from the Yale Centre for Natural Carbon Capture.**Data and materials availability:** All data are available in the main text or the supplementary materials. The weathering model was developed in MATLAB v.R2019a, data processing in both MATLAB v.R2019a and Python v.3.7. MATLAB and Python codes developed for this study belong to the Leverhulme Centre for Climate Change Mitigation. The authors will make these codes and the modified codes in CLM5 developed in this study available upon reasonable request. All other data required to evaluate the conclusions are in the main paper or Supplementary Materials.

## SUPPLEMENTARY MATERIALS
Tables S1 to S4
Figs S1 to S22
Data Files D1 to D4

**References**
1. Executive Office of the President (2021) The long-term strategy of the United States: pathways to net-zero greenhouse gas emissions by 2050 (US Department of State).
2. E. Larson *et al.* (2020) Net-Zero America: potential pathways, infrastructure, and impacts, interim report. (Princeton University, Princeton, NJ), p 345 pp.
3. J. H. Williams *et al.*, Carbon-neutral pathways for the United States. *AGU Advances* **2**, e2020AV000284 (2021).
4. J. Hansen *et al.*, Young people's burden: requirement of negative $CO_2$ emissions. *Earth Syst. Dynam.* **8**, 577–616 (2017).
5. J. Hartmann *et al.*, Enhanced chemical weathering as a geoengineering strategy to reduce atmospheric carbon dioxide, supply nutrients, and mitigate ocean acidification. *Review of Geophysics* **51**, 113–149 (2013).
6. D. J. Beerling *et al.*, Potential for large-scale $CO_2$ removal via enhanced rock weathering with croplands. *Nature* **583**, 242–248 (2020).
7. I. B. Kantola, M. D. Masters, D. J. Beerling, S. P. Long, E. H. DeLucia, Potential of global croplands and bioenergy crops for climate change mitigation through deployment for enhanced weathering. *Biology Letters* **13**, 20160714 (2017).
8. D. J. Beerling *et al.*, Farming with crops and rocks to address global climate, food and soil security. *Nature Plants* **4**, 138–147 (2018).
9. P. Köhler, J. Hartmann, D. A. Wolf-Gladrow, Geoengineering potential of artificially enhanced silicate weathering of olivine. *Proceedings of the National Academy of Sciences* **107**, 20228–20233 (2010).
10. L. L. Taylor *et al.*, Enhanced weathering strategies for stabilizing climate and averting ocean acidification. *Nature Climate Change* **6**, 402–406 (2016).26

11. N. Vakilifard, E. u. P. Kantzas, N. R. Edwards, P. B. Holden, D. J. Beerling, The role of enhanced rock weathering deployment with agriculture in limiting future warming and protecting coral reefs. *Environmental Research Letters* **16**, 094005 (2021).
12. The White House (2022) Justice40: A whole-of-government initiative. (The White House, 1600 Pennsylvania Ave NW, Washington DC 20500).
13. D. J. Beerling *et al.*, Enhanced weathering in the U.S. Corn Belt delivers carbon removal with agronomic benefits. arXiv [Preprint]. https://arxiv.org/abs/2307.05343 (accessed 8 August 2023) (2023).
14. National Academies of Science Engineering and Medicine (2019) Negative research technologies and reliable sequestration: a research agenda. (The National Academies Press, Washington DC).
15. House of Representatives (117th Congress 2021–2022) Agriculture, Rural Development, Food and Drug Administration, and Related Agencies Appropriations Bill, 2023.
16. E. P. Kantzas *et al.*, Substantial carbon drawdown potential from enhanced rock weathering in the United Kingdom. *Nature Geoscience* **15**, 382–389 (2022).
17. P. B. Holden *et al.*, Climate–carbon cycle uncertainties and the Paris Agreement. *Nature Climate Change* **8**, 609–613 (2018).
18. J.-F. Mercure *et al.*, Environmental impact assessment for climate change policy with the simulation-based integrated assessment model E3ME-FTT-GENIE. *Energy strategy reviews* **20**, 195–208 (2018).
19. IPCC, *Global warming of 1.5 °C. An IPCC special report on the impacts of global warming of 1.5 °C above pre-industrial levels and related global greenhouse gas emission pathways, in the context of strengthening the global response to the threat of climate change, sustainable developemnet, and efforts to eradicate poverty* (2018), vol. 1, pp. 43–50.
20. J. Grafström, R. Poudineh (2021) A critical assessment of learning curves for solar and wind power technologies. OIES Paper: EL No. 43. (The Oxford Institute for Energy Studies, Oxford).
21. E. Baik *et al.*, Geospatial analysis of near-term potential for carbon-negative bioenergy in the United States. *Proceedings of the National Academy of Sciences* **115**, 3290–3295 (2018).
22. G. M. Domke, S. N. Oswalt, B. F. Walters, R. S. Morin, Tree planting has the potential to increase carbon sequestration capacity of forests in the United States. *Proceedings of the National Academy of Sciences* **117**, 24649–24651 (2020).
23. R. M. Eufrasio *et al.*, Environmental and health impacts of atmospheric $CO_2$ removal by enhanced rock weathering depend on nations' energy mix. *Communications Earth & Environment* **3**, 106 (2022).
24. A. Somani, T. K. Nandi, S. K. Pal, A. K. Majumder, Pre-treatment of rocks prior to comminution – A critical review of present practices. *International journal of mining science and technology* **27**, 339–348 (2017).
25. S. Zhang *et al.*, River chemistry constraints on the carbon capture potential of surficial enhanced rock weathering. *Limnology and Oceanography* **67**, S148–S157 (2022).
26. J. Rassmann, B. Lansard, L. Pozzato, C. Rabouille, Carbonate chemistry in sediment porewaters of the Rhône River delta driven by early diagenesis (northwestern Mediterranean). *Biogeosciences* **13**, 5379–5394 (2016).
27. Y. Kanzaki, N. J. Planavsky, C. T. Reinhard, New estimates of the storage permanence and ocean co-benefits of enhanced rock weathering. *PNAS Nexus* **2**, pgad059 (2023).
28. A. E. Russell, D. A. Laird, A. P. Mallarino, Nitrogen fertilization and cropping system impacts on soil quality in midwestern mollisols. *Soil Science Society of America Journal* **70**, 249–255 (2006).
29. L. A. Alves *et al.*, Biological $N_2$ fixation by soybeans grown with or without liming on acid soils in a no-till integrated crop-livestock system. *Soil and Tillage Research* **209**, 104923 (2021).
30. T. O. West, A. C. McBride, The contribution of agricultural lime to carbon dioxide emissions in the United States: dissolution, transport, and net emissions. *Agriculture, Ecosystems & Environment* **108**, 145–154 (2005).
31. J. Baffes, W. C. Koh (2022) Fertilizer prices expected to remain higher for longer. in *World Bank Blogs*.

32. C. Nevison, E. Holland, A reexamination of the impact of anthropogenically fixed nitrogen on atmospheric $N_2O$ and the stratospheric $O_3$ layer. *Journal of Geophysical Research: Atmospheres* **102**, 25519–25536 (1997).
33. E. Blanc-Betes *et al.*, In silico assessment of the potential of basalt amendments to reduce $N_2O$ emissions from bioenergy crops. *Global Change Biology: Bioenergy* **13**, 224–241 (2020).
34. M. Val Martin *et al.*, Improving nitrogen cycling in a land surface model (CLM5) to quantify soil $N_2O$, NO and $NH_3$ emissions from enhanced rock weathering with croplands. *Geoscientific Model Development Discuss* **2023**, 1–32 (2023).
35. J. A. Logan, Nitrogen oxides in the troposphere: Global and regional budgets. *Journal of Geophysical Research: Oceans* **88**, 10785–10807 (1983).
36. J. M. McGrath *et al.*, An analysis of ozone damage to historical maize and soybean yields in the United States. *Proceedings of the National Academy of Sciences* **112**, 14390–14395 (2015).
37. S. Avnery, D. L. Mauzerall, A. M. Fiore, Increasing global agricultural production by reducing ozone damages via methane emission controls and ozone-resistant cultivar selection. *Global Change Biology* **19**, 1285–1299 (2013).
38. M. Val Martin *et al.*, How emissions, climate, and land use change will impact mid-century air quality over the United States: a focus on effects at national parks. *Atmospheric Chemistry and Physics* **15**, 2805–2823 (2015).
39. M. S. Mkhabela, R. Gordon, D. Burton, A. Madani, W. Hart, Effect of lime, dicyandiamide and soil water content on ammonia and nitrous oxide emissions following application of liquid hog manure to a marshland soil. *Plant and Soil* **284**, 351–361 (2006).
40. J. Chen *et al.*, Seasonal modeling of $PM_{2.5}$ in California's San Joaquin Valley. *Atmospheric Environment* **92**, 182–190 (2014).
41. A. C. Dalmora *et al.*, Nanoparticulate mineral matter from basalt dust wastes. *Chemosphere* **144**, 2013–2017 (2016).
42. A. Protopapas, C. J. Kruse, L. E. Olson, Modal comparison of domestic freight transportation effects on the general public. *Transportation research record* **2330**, 55–62 (2013).
43. S. Fuss *et al.*, Negative emissions—Part 2: Costs, potentials and side effects. *Environmental Research Letters* **13**, 063002 (2018).
44. B. Guenet *et al.*, Can $N_2O$ emissions offset the benefits from soil organic carbon storage? *Global Change Biology* **27**, 237–256 (2021).
45. Y. Rui *et al.*, Reply to Lajtha and Silva: Agriculture and soil carbon persistence of grassland-derived Mollisols. *Proceedings of the National Academy of Sciences* **119**, e2204142119 (2022).
46. C. D. Kolstad (2022) What Is killing the US coal industry?  (Stanford institute for economic policy research), p 2000.
47. Office of Fossil Energy and Carbon Management (2021) The Infrastructure Investment and Jobs Act: Opportunities to Accelerate Deployment in Fossil Energy and Carbon Management Activities. ed U. D. o. Energy.
48. K. Rennert *et al.*, Comprehensive evidence implies a higher social cost of $CO_2$. *Nature* **610**, 687–692 (2022).
49. K. Moffat, J. Lacey, A. Zhang, S. Leipold, The social licence to operate: a critical review. *Forestry: An International Journal of Forest Research* **89**, 477–488 (2016).
50. B. K. Sovacool, C. M. Baum, S. Low, Reviewing the sociotechnical dynamics of carbon removal. *Joule* **7**, 57–82 (2023).
51. V. Campbell-Arvai, P. S. Hart, K. T. Raimi, K. S. Wolske, The influence of learning about carbon dioxide removal (CDR) on support for mitigation policies. *Climatic Change* **143**, 321–336 (2017).
52. E. Spence, E. Cox, N. Pidgeon, Exploring cross-national public support for the use of enhanced weathering as a land-based carbon dioxide removal strategy. *Climatic Change* **165**, 23 (2021).
53. E. Cox, E. Spence, N. Pidgeon, Public perceptions of carbon dioxide removal in the United States and the United Kingdom. *Nature Climate Change* **10**, 744–749 (2020).
54. S. K. Sweet, J. P. Schuldt, J. Lehmann, D. A. Bossio, D. Woolf, Perceptions of naturalness predict US public support for Soil Carbon Storage as a climate solution. *Climatic Change* **166**, 22 (2021).

55. N. Pidgeon, C. C. Demski, From nuclear to renewable: Energy system transformation and public attitudes. *Bulletin of the Atomic Scientists* **68**, 41–51 (2012).
56. C. Blattner, Just transition for agriculture? A critical step in tackling climate change. *Journal of Agriculture, Food Systems, and Community Development* **9**, 53–58 (2020).
57. Leverhulme Centre for Climate Change Mitigation (2023) Accelerating global commercialisation of ERW for atmospheric carbon removal. (https://www.sheffield.ac.uk/lc3m/about/accelerating-global-commercialisation-erw-atmospheric-carbon-removal, accessed 8 August 2023).
58. P. Horton, S. P. Long, P. Smith, S. A. Banwart, D. J. Beerling, Technologies to deliver food and climate security through agriculture. *Nature Plants* **7**, 250–255 (2021).
59. PetDB database (2021) EarthChem Syntheses. (https://earthchem.org/data-access/earthchem-syntheses/ accessed 21 September 2021).
60. GEOROC database (2022) Geochemistry of rocks of the oceans and continents. (https://georoc.eu/, accessed 29 June 2022).
61. K. Lehnert, Y. Su, C. H. Langmuir, B. Sarbas, U. Nohl, A global geochemical database structure for rocks. *Geochemistry Geophysics Geosystems* **1**, 1012 (2000).
62. K. Hollocher (2004) CIPW Norm calculation program. . (www.whitman.edu, Union College, Schenectady, NY 12308).
63. A. Johannsen, *A descriptive petrography of the igneous rocks: the intermediate rocks* (University of Chicago Press, Chicago, 1931), vol. 1, pp. 267.
64. K. Hollocher, Calculation of a CIPW norm from a bulk chemical analysis.
65. W. Cross, J. P. Iddings, L. Pirsson, H. Washington, Modifications of the" Quantitative System of Classification of Igneous Rocks". *The Journal of Geology* **20**, 550–561 (1912).
66. R. Le Maitre, Some problems of the projection of chemical data into mineralogical classifications. *Contributions to Mineralogy and Petrology* **56**, 181–189 (1976).
67. H. Nesbitt, G. Young, Early Proterozoic climates and plate motions inferred from major element chemistry of lutites. *Nature* **299**, 715–717 (1982).
68. M. Le Bas, A. L. Streckeisen, The IUGS systematics of igneous rocks. *Journal of the Geological Society* **148**, 825–833 (1991).
69. S. L. Brantley, Kinetics of Mineral Dissolution "Kinetics of Mineral Dissolution" in Kinetics of Water-Rock Interaction*,* S. L. Brantley, J. D. Kubicki, A. F. White, Eds. (Springer, New York, 2008), 10.1007/978-0-387-73563-4_5 chap. 5, pp. 151–210.
70. National Minerals Information Center (2021, accessed April 2023) USGS aggregates time series data by state, type, and end use. ( https://www.usgs.gov/media/files/usgs-aggregates-time-series-data-state-type-and-end-use).
71. G. Danabasoglu *et al.*, The Community Earth System Model Version 2 (CESM2). *Journal of Advances in Modeling Earth Systems* **12**, e2019MS001916 (2020).
72. D. M. Lawrence *et al.*, The Community Land Model Version 5: Description of New Features, Benchmarking, and Impact of Forcing Uncertainty. *Journal of Advances in Modeling Earth Systems* **11**, 4245–4287 (2019).
73. W. R. Wieder *et al.*, Beyond static benchmarking: using experimental manipulations to evaluate land model assumptions. *Global Biogeochemical Cycles* **33**, 1289–1309 (2019).
74. D. L. Lombardozzi *et al.*, Simulating agriculture in the Community Land Model Version 5. *Journal of Geophysical Research: Biogeosciences* **125**, e2019JG005529 (2020).
75. D. M. Lawrence *et al.*, The Land Use Model Intercomparison Project (LUMIP) contribution to CMIP6: rationale and experimental design. *Geoscientific Model Development* **9**, 2973–2998 (2016).
76. USDA (2019) Fertilizer use and price. in *Economic Research Services* (US Department of Agriculture).
77. FAO (2019) World fertilizer trends and outlook to 2022. (Rome).
78. C. Nevison (2021) Indirect $N_2O$ emissions from nitrogen used in agriculture. in *IPCC Good Practice Guidance and Uncertainty Management in National Greenhouse Gas Inventories* (Intergovernmental Panel on Climate Change).

79. C. Lu *et al.*, Century-long changes and drivers of soil nitrous oxide ($N_2O$) emissions across the contiguous United States. *Global Change Biology* **28**, 2505–2524 (2022).
80. L. K. Emmons *et al.*, The chemistry mechanism in the Community Earth System Model Version 2 (CESM2). *Journal of Advances in Modeling Earth Systems* **12**, e2019MS001882 (2020).
81. X. Liu *et al.*, Description and evaluation of a new four-mode version of the Modal Aerosol Module (MAM4) within version 5.3 of the Community Atmosphere Model. *Geoscientific Model Development* **9**, 505–522 (2016).
82. Z. Lu *et al.*, Radiative forcing of nitrate aerosols from 1975 to 2010 as simulated by MOSAIC module in CESM2-MAM4. *Journal of Geophysical Research: Atmospheres* **126**, e2021JD034809 (2021).
83. K. Riahi *et al.*, The Shared Socioeconomic Pathways and their energy, land use, and greenhouse gas emissions implications: An overview. *Global environmental change* **42**, 153–168 (2017).
84. R. Gelaro *et al.*, The Modern-Era Retrospective Analysis for Research and Applications, Version 2 (MERRA-2). *Journal of Climate* **30**, 5419–5454 (2017).
85. J. F. Lamarque *et al.*, CAM-chem: description and evaluation of interactive atmospheric chemistry in the Community Earth System Model. *Geoscientific Model Development* **5**, 369–411 (2012).
86. R. H. Schwantes *et al.*, Evaluating the impact of chemical complexity and horizontal resolution on tropospheric ozone over the conterminous US with a global variable resolution chemistry model. *Journal of Advances in Modeling Earth Systems* **14**, e2021MS002889 (2022).
87. R. Van Dingenen *et al.*, The global impact of ozone on agricultural crop yields under current and future air quality legislation. *Atmospheric Environment* **43**, 604–618 (2009).
88. USDA (2021, accessed 21 April 2023) Statistics by state. (US Department of Agriculture).
89. FAO (2018, accessed 21 April 2023) Food and agriculture 2050 data portal version 2.0 BETA. (FAO).
90. J. L. Palandri, Y. K. Kharaka, A compilation of rate parameters of water-mineral interaction kinetics for application to geochemical modeling. *Open File Report 2004-1068*, 64 pages (2004).
91. S. A. Banwart, A. Berg, D. J. Beerling, Process-based modeling of silicate mineral weathering responses to increasing atmospheric $CO_2$ and climate change. *Global Biogeochemical Cycles* **23**, GB4013 (2009).
92. P. N. Nelson, N. Su, Soil pH buffering capacity: a descriptive function and its application to some acidic tropical soils *Soil Research* **48**, 201–207 (2010).
93. T. E. Cerling, Carbon dioxide in the atmosphere; evidence from Cenozoic and Mesozoic Paleosols. *American Journal of Science* **291**, 377–400 (1991).
94. R. B. Jackson *et al.*, A global analysis of root distributions for terrestrial biomes. *Oecologia* **108**, 389–411 (1996).
95. T. Ozbun (2022) Brazil: sugar cane yield 2010. (https://www.statista.com/statistics/742523/sugar-cane-yield-brazil/).
96. E. A. Heaton, F. G. Dohleman, S. P. Long, Meeting US biofuel goals with less land: the potential of Miscanthus. **14**, 2000–2014 (2008).
97. P. Renforth, The potential of enhanced weathering in the UK. *International Journal of Greenhouse Gas Control* **10**, 229–243 (2012).
98. T. J. Brown *et al.* (2010) Underground mining of aggregates. Main report. (Camborne School of Mines, Penryn, UK), p 322 pp.
99. S. Roussanaly *et al.*, Offshore power generation with carbon capture and storage to decarbonise mainland electricity and offshore oil and gas installations: A techno-economic analysis. *Applied Energy* **233**, 478–494 (2019).
100. Indeed (accessed April 2023) (www.indeed.com).
101. AUMA (accessed April 2023) Chemical Engineering. (https://www.chemengonline.com/2020-annual-cepci-average-value/).
102. EIA (accessed April 2023) Electricity Data Browser. (US Energy Information Administration, https://www.eia.gov/electricity/data/browser/).

103. D. R. Woods, *Rules of thumb in engineering practice* (Wiley-VCH Verlag GmbH & Col KG2A, Weinheim, 2007).
104. K. Gerdes, W. M. Summers, J. Wimer (2011) Quality Guidelines for Energy System Studies: Cost Estimation Methodology for NETL Assessments of Power Plant Performance. (National Energy Technology Laboratory (NETL), Pittsburgh, PA, United States).
105. S. Morrell, Predicting the overall specific energy requirement of crushing, high pressure grinding roll and tumbling mill circuits. *Minerals engineering* **22**, 544–549 (2009).
106. J. F. Mercure *et al.*, Macroeconomic impact of stranded fossil fuel assets. *Nature Climate Change* **8**, 588–593 (2018).
107. A. A. Phadke, A. Khandekar, N. Abhyankar, D. Wooley, D. Rajagopal (2021) Why regional and long-haul trucks are primed for electrification now. in *Energy Technologies Area, Publications. Berkeley Lab*.
108. H. Liimatainen, O. van Vliet, D. Aplyn, The potential of electric trucks – An international commodity-level analysis. *Applied Energy* **236**, 804-814 (2019).
109. AEO2020 (2020) Annual Energy Outlook 2020 with projections to 2050. (U.S. Energy Information Administration (EIA)).
110. BTS.GOV, Table 4-17: Class I Rail Freight Fuel Consumption and Travel. https://www.bts.gov/archive/publications/national_transportation_statistics/table_04_17.
111. IEA (2019) The Future of Rail, Opportunities for energy and the environment. International Energy Agency.
112. SMMT (2010) Truck specification for best operational efficiency, guide, department for transport.
113. MDS (2012) Analysis of road and rail costs between coal mines and power stations, MDS TRANSMODAL LIMITED.
114. Oscar Delgado, Felipe Rodríguez, R. Muncrief (2017) Fuel efficiency technology in European heavy-duty vehicles: baseline and potential for the 2020–2030 time frame. The International Council of Clean Transportation. in *White paper*.
115. J. Hartmann, N. Moosdorf, The new global lithological map database GLiM: A representation of rock properties at the Earth surface. *Geochemistry, Geophysics, Geosystems* **13** (2012).
116. D. J. Maguire, "ArcGIS: general purpose GIS software system" in Encyclopedia of GIS, S. Shekhar, H. Xiong, Eds. (Springer US, Boston, MA, 2008), 10.1007/978-0-387-35973-1_68, pp. 25-31.
117. M. J. Sprung *et al.* (2018, accessed April 2023) Freight facts and figures. ed U. D. o. Transportation (US Department of Transportation, https://www.bts.dot.gov/sites/bts.dot.gov/files/docs/FFF_2017_Full_June2018revision.pdf).
118. USGS (2016, accessed April 2021) Water data for the nation. (United States Geological Survey, http://waterdata.usgs.gov/nwis/).
119. GRASS Development Team (2017) Geographic Resources Analysis Support System (GRASS GIS). (Open Source Geospatial Foundation).
120. B. Lehner, K. Verdin, A. Jarvis, New global hydrography derived from spaceborne elevation data. *Transactions of the American Geophysical Union* **89**, 93–94 (2008).
121. R. E. Zeebe, D. Wolf-Gladrow, *CO2 in seawater: equilibrium, kinetics, isotopes* (Gulf Professional Publishing, 2001).
122. J.-P. Gattuso *et al.* (2015) Seawater carbonate chemistry. (http://cran.r-project.org/package=seacarb).
123. R. Marsh, S. Müller, A. Yool, N. Edwards, Incorporation of the C-GOLDSTEIN efficient climate model into the GENIE framework: "eb_go_gs" configurations of GENIE. *Geoscientific Model Development* **4**, 957–992 (2011).
124. N. R. Edwards, R. Marsh, Uncertainties due to transport-parameter sensitivity in an efficient 3-D ocean-climate model. *Climate Dynamics* **24**, 415–433 (2005).
125. A. Ridgwell, J. C. Hargreaves, Regulation of atmospheric $CO_2$ by deep-sea sediments in an Earth system model. *Global Biogeochemical Cycles* **21**, GB2008 (2007).
126. A. Ridgwell, D. N. Schmidt, Past constraints on the vulnerability of marine calcifiers to massive carbon dioxide release. *Nature Geoscience* **3**, 196–200 (2010).

127. C. T. Reinhard *et al.*, Oceanic and atmospheric methane cycling in the cGENIE Earth system model – release v0.9.14. *Geoscientific Model Development* **13**, 5687–5706 (2020).
128. A. Ridgwell *et al.*, Marine geochemical data assimilation in an efficient Earth System Model of global biogeochemical cycling. *Biogeosciences* **4**, 87–104 (2007).
129. M. Meinshausen *et al.*, The RCP greenhouse gas concentrations and their extensions from 1765 to 2300. *Climatic Change* **109**, 213 (2011).
130. Cambridge Econometrics (2022) E3ME Manual: Version 9.0. (https://www.e3me.com/what/e3me/).

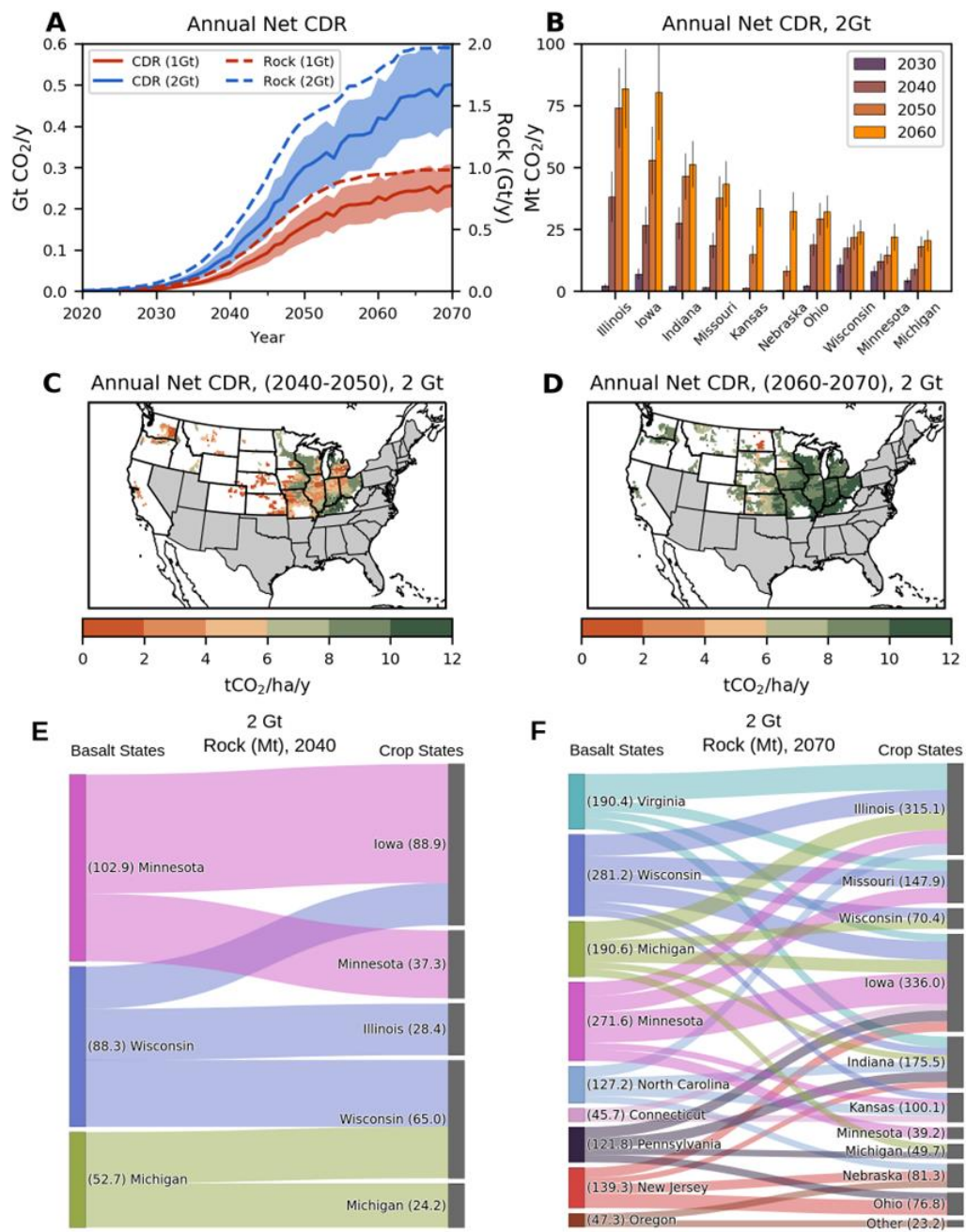

**Fig. 1. Atmospheric carbon dioxide removal by enhanced weathering with United States agriculture.**
(**A**) Net annual cumulative carbon dioxide removal (CDR) by EW as constrained by 1 and 2 Gt rock yr$^{-1}$ scenarios, 2020-2070. Shaded area shows the 90% uncertainty envelope due to differences in the mineralogy of basalt sourced from the supply states. (**B**) Mean (with 90% confidence limits) annual CDR rates of the top ten states (2 Gt rock yr$^{-1}$ scenario). (**C**) Spatial patterns of net annual CDR rates per hectare in 2040-2050 and (**D**) in 2060-2070 for the 2 Gt rock yr$^{-1}$ scenario. (**E**) and (**F**) Sankey diagrams illustrating the main transfer pathways of crushed rock from basalt source states to recipient cropland states in 2040 and 2070 respectively, for the 2 Gt yr$^{-1}$ rock scenario; only fluxes >20 Mt yr$^{-1}$ are shown for clarity.

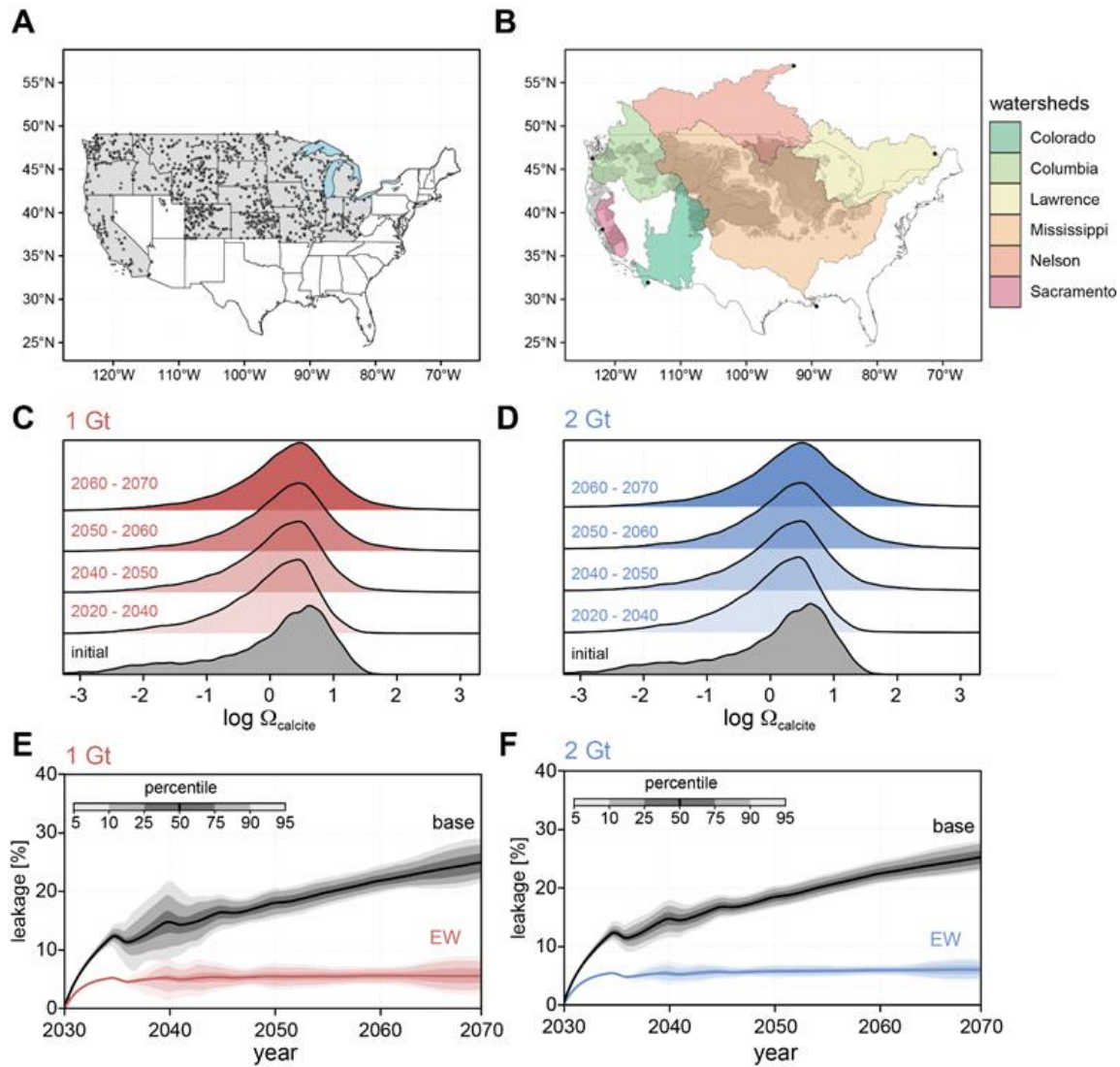

**Fig. 2. River and ocean responses to enhanced rock weathering.**
(**A**) Locations of river/stream sites from the aqueous geochemical database used to estimate river calcium carbonate saturation states ($\Omega_{calcite}$). Filled circles show individual monitoring sites, grey shading indicates the U.S. states in which EW was applied. (**B**) Watersheds over which river cation and dissolved inorganic carbon data were interpolated for use in our Earth system model. Shaded polygons show watershed extent; filled circles show outflow locations. The six large watersheds considered are the Mississippi, Colorado, Columbia, Sacramento, Lawrence, and Nelson. (**C**) Frequency distributions of $\Omega$ for U.S. river systems in the background state (grey) and for each decade between 2020-2070 in the 1 Gt yr$^{-1}$ rock and (**D**) 2 Gt yr$^{-1}$ rock scenarios. (**E**) Carbon leakage from the ocean during EW for 1 Gt yr$^{-1}$ rock and (**F**) the 2 Gt yr$^{-1}$ rock scenarios. Solid lines and shaded regions show the median and percentile values, respectively, for our 984-member model ensemble. Base leakage refers to CO$_2$ outgassing from the ocean; EW is the residual re-release of CO$_2$ captured through EW (see *17*).

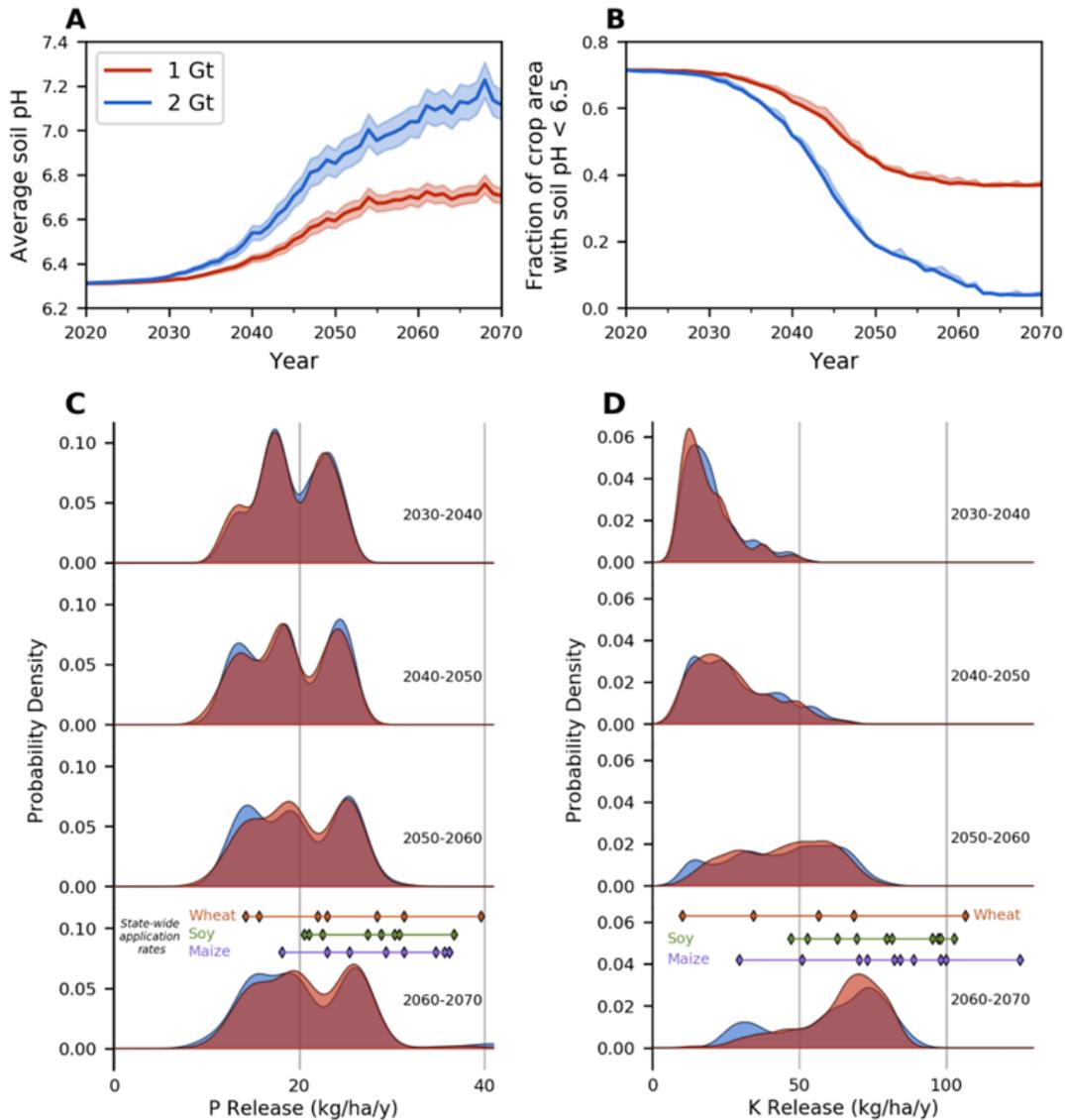

**Fig. 3. Benefits of enhanced weathering for agricultural soils.**
(**A**) Average topsoil (0-15 cm depth) pH for U.S. agricultural lands, and top ten Corn Belt states by CDR potential, with EW deployment for the 1 and 2 Gt yr$^{-1}$ rock scenarios. (**B**) Decreasing fraction of acidified lands over time with EW deployment for the 1 and 2 Gt yr$^{-1}$ rock scenarios. (**C**) Frequency histograms of phosphorus (P) release and (**D**) potassium (K) release by EW with basalt over successive decades for Corn Belt states (2030-2070). Also indicated in (C) and (D) is the application rates of P and K fertilizers for states growing soybean, maize and wheat (see *17*). Shading in (A) and (B) denotes 90% confidence limits.

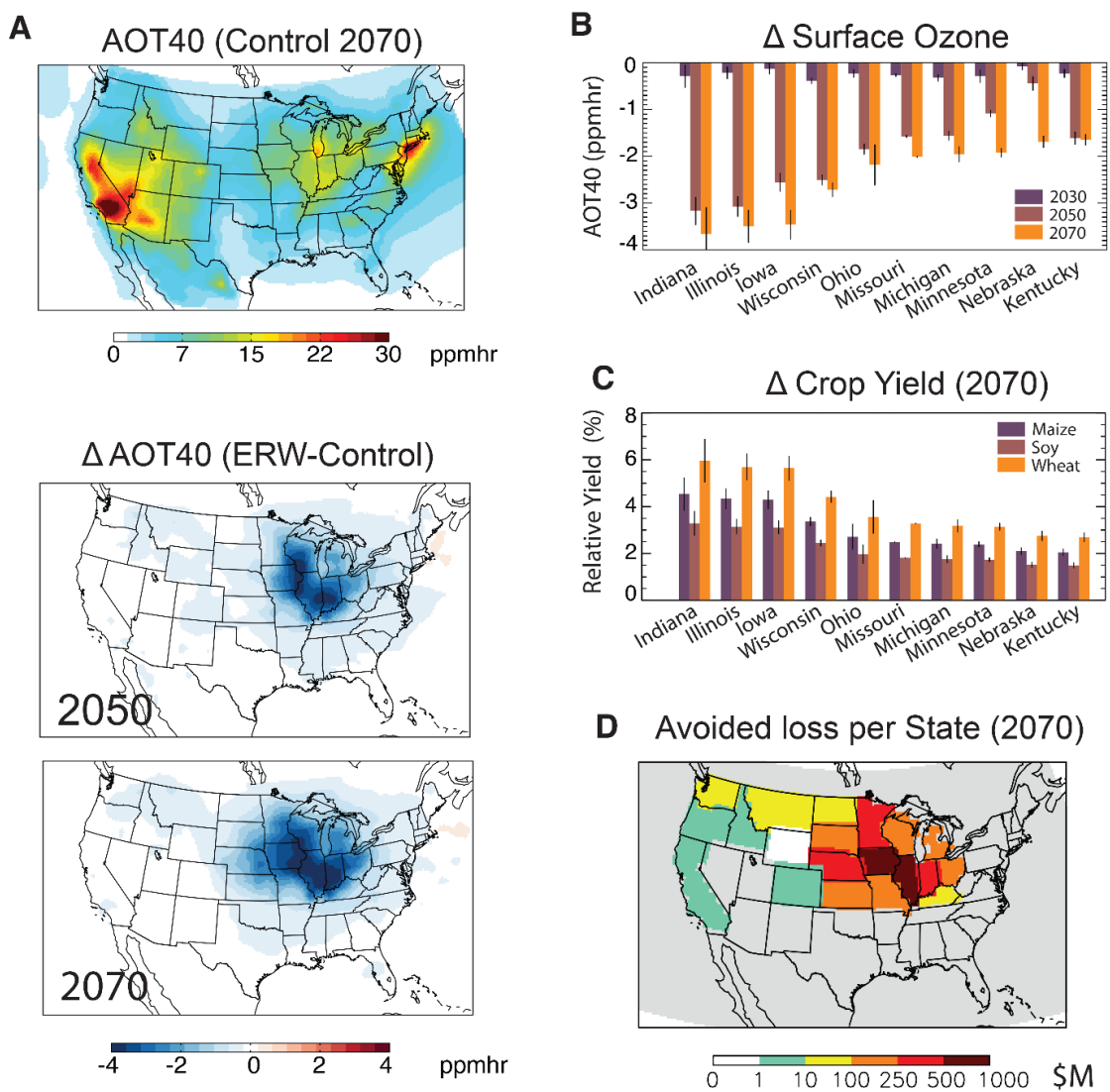

**Fig. 4. Benefits of enhanced weathering for surface ozone and crop production.**
(**A**) Simulated AOT40 levels for 2070 (control; anthropogenic emissions + biomass burning + biogenic emissions, no EW effects), defined as Accumulated dose of Ozone over a Threshold of 40 ppb, with widespread reductions by 2050 and 2070 due to EW lowering soil nitrogen trace gas emissions. (**B**) Average change in AOT40 (± 1 s.e.) for ten states with the largest reductions. (**C**) Corresponding calculated increases in crop yields (soybean, maize and wheat) for 2070. (**D**) Calculated avoided economic yield losses per state from lower AOT40 levels by 2070 due to improved air quality. Reported is the total for current areas of maize, soybean and wheat in each state.

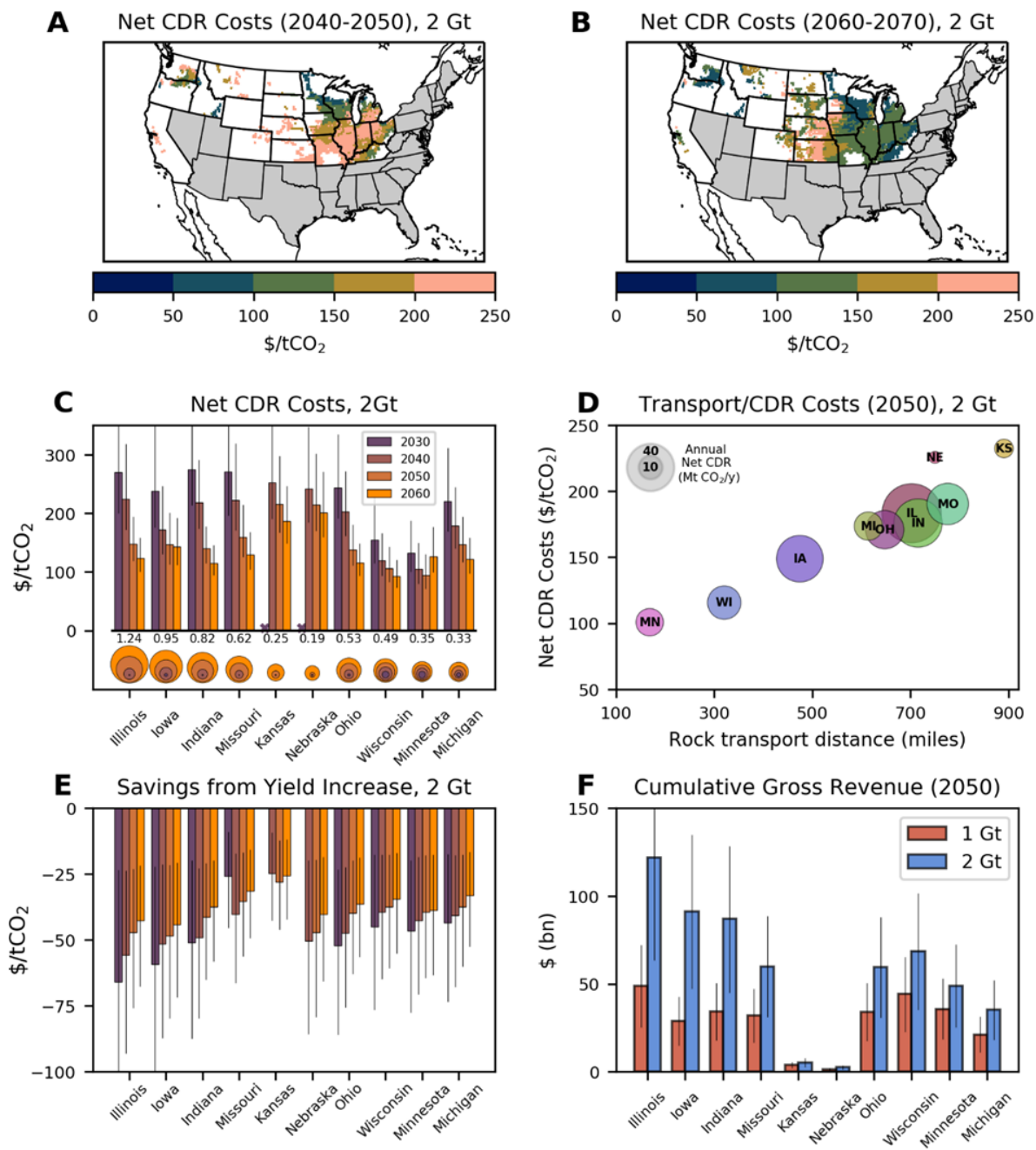

**Fig. 5. Costs and revenue of enhanced weathering implementation in the U.S.**
(**A**) Geospatial pattern costs of carbon dioxide removal (CDR) by EW for the 2 Gt yr$^{-1}$ rock scenario in 2040-2050 and (**B**) 2060-2070. (**C**) Averaged state-level costs of CDR by EW for successive decades between 2030 and 2060. Circles show corresponding cumulative CDR by decade for each state. (**D**) Relationship between state-specific costs of CDR and average distance between basalt supply state and farmland for 2060; size of size indicate cumulative CDR in 2060. (**E**) Discount CDR costs calculated from yield gains. (**F**) Cumulative revenue by 2050 of the top ten states for 1 and 2 Gt rock yr$^{-1}$ scenarios, as calculated with a social cost of carbon ($185 tCO$_2^{-1}$) (*48*). Errors in all plots are the 90% confidence limits.